\documentclass[conference,compsoc]{IEEEtran}

\usepackage{cite}
\usepackage[pdftex]{graphicx}
\graphicspath{{./figures/}}
\DeclareGraphicsExtensions{.pdf,.jpeg,.png,.jpg}
\usepackage{array}
\usepackage[caption=false,font=footnotesize,labelfont=sf,textfont=sf]{subfig}
\usepackage[hyphens]{url}

\usepackage{xcolor}
\usepackage[normalem]{ulem}
\usepackage{hyperref}
\hypersetup{
  pdfborderstyle={/S/U/W 0.6},
  urlbordercolor=[rgb]{0.35,0.50,0.60},
}
\usepackage{bookmark}

\usepackage{amsmath,amsfonts,amsthm,amssymb}
\usepackage{aliascnt}

\usepackage[linesnumbered,ruled,vlined]{algorithm2e}
\SetKwInput{KwData}{Input}
\SetKwRepeat{Do}{do}{while}
\usepackage{enumitem}
\usepackage[operators,sets]{cryptocode}
\usepackage{dashbox}
\createprocedureblock{procb}{center,boxed}{}{}{}
\usepackage{booktabs} 

\usepackage{xfrac}

\usepackage{tikz}
\usepackage{hwemoji}
\usepackage{upgreek}
\usepackage{pifont}

\SetCommentSty{mycommfont}

\definecolor{codegreen}{rgb}{0,0.6,0}
\definecolor{codegray}{rgb}{0.5,0.5,0.5}
\definecolor{codepurple}{rgb}{0.58,0,0.82}
\definecolor{backcolour}{rgb}{0.95,0.95,0.92}

\usepackage{listings}
\theoremstyle{plain}

\newaliascnt{proposition}{theorem}
\newtheorem{proposition}[proposition]{Proposition}
\aliascntresetthe{proposition}
\newaliascnt{lemma}{theorem}

\aliascntresetthe{lemma}
\newtheorem{definition}{Definition}[section]
\newaliascnt{remark}{theorem}
\newtheorem{remark}[remark]{{\bf Remark}}
\aliascntresetthe{remark}

\def\sectionautorefname{Section}
\def\subsectionautorefname{Section}

\def\subsectionautorefname{Section}

\newcommand{\eps}{\varepsilon}
\newcommand{\del}{\delta}

\newcommand{\M}{\mathcal{M}}
\newcommand{\A}{A}
\newcommand{\x}{x}
\newcommand{\X}{\mathcal{X}}
\newcommand{\Y}{\mathcal{Y}}
\newcommand{\R}{\mathbb{R}}
\newcommand{\Rej}{\mathcal{R}}
\newcommand{\sig}{\gamma}
\newcommand{\OO}{\Omega}

\newcommand{\eqdef}{:=}

\newcommand{\fpr}{\alpha}
\newcommand{\fnr}{\beta}
\newcommand{\fp}{\mathrm{FP}}
\newcommand{\fn}{\mathrm{FN}}
\newcommand{\tp}{\mathrm{TP}}
\newcommand{\tn}{\mathrm{TN}}

\newcommand{\Fclass}{\mathcal{F}}

\newcommand{\Beta}{\mathrm{Beta}}
\newcommand{\Bin}{\mathrm{Binomial}}

\newcommand{\expec}[2]{\underset{#1}{\mathbb{E}}\left[#2\right]}
\newcommand{\prob}[2]{\underset{#1}{\mathbb{P}}\left[#2\right]}

\newcommand{\lap}{\mathrm{Lap}}
\newcommand{\gaus}{\mathrm{Gauss}}
\newcommand{\enc}{\mathsf{Enc}}
\newcommand{\dec}{\mathsf{Dec}}
\newcommand{\pk}{\mathsf{pk}}
\newcommand{\sk}{\mathsf{sk}}
\newcommand{\msg}{\mathsf{msg}}
\newcommand{\kgen}{\mathsf{KGen}}
\newcommand{\hash}{\mathsf{Hash}}
\newcommand{\key}{\mathsf{k}}

\newcommand{\symohe}{\ensuremath{\mathsf{symOHE}}}
\newcommand{\clientcms}{\ensuremath{\mathsf{client}\text{-}\mathsf{CMS}}}
\newcommand{\clienthcms}{\ensuremath{\mathsf{client}\text{-}\mathsf{HCMS}}}
\newcommand{\secretsharesfn}{\ensuremath{\mathsf{SecretShares}}}
\newcommand{\snipsfn}{\ensuremath{\mathsf{SNIPs}}}
\newcommand{\verifyfn}{\ensuremath{\mathsf{Verify}}}
\newcommand{\seedvar}{\ensuremath{\mathsf{seed}}}
\newcommand{\seedgenfn}{\ensuremath{\mathsf{SGen}}}
\newcommand{\snipfn}{\ensuremath{\mathsf{SNIP}}}
\newcommand{\prngfn}{\ensuremath{\mathsf{PRNG}}}
\newcommand{\shafn}{\ensuremath{\mathsf{SHA256}}}
\newcommand{\aesfn}{\ensuremath{\mathsf{AES128}}}
\newcommand{\ohefn}{\ensuremath{\mathsf{OHE}}}
\newcommand{\shareonefield}{\ensuremath{\mathsf{share1}}}
\newcommand{\sharetwofield}{\ensuremath{\mathsf{share2}}}

\newcommand{\hdr}[2]{\href{https://anonymous.4open.science/r/ios17-dyld-headers-DifferentialPrivacy-E187/#1}{#2}}

\newif\ifshowchanges
\showchangesfalse
\definecolor{changeadd}{rgb}{0.05,0.35,0.75}
\definecolor{changedelete}{rgb}{0.72,0.14,0.12}
\newcommand{\chgprotect}[1]{\mbox{#1}}
\newcommand{\chgadd}[1]{\ifshowchanges{\begingroup\color{changeadd}#1\endgroup}\else#1\fi}
\newcommand{\chgdel}[1]{\ifshowchanges{\begingroup\color{changedelete}\sout{#1}\endgroup}\else\ignorespaces\fi}
\newcommand{\chgrep}[2]{\ifshowchanges\chgdel{#1}\chgadd{#2}\else#2\fi}
\newcommand{\chgdeletefigure}[3]{%
  \ifshowchanges
    \begin{figure}[!h]
      \centering
      \setlength{\fboxsep}{6pt}%
      \fcolorbox{changedelete}{white}{%
        \parbox{0.88\linewidth}{\centering\small\textcolor{changedelete}{Removed in camera-ready: #2}}%
      }
      \caption{\chgdel{#3}}
      \label{#1}
    \end{figure}
  \fi
}

\newcommand{\cmark}{\ding{51}}%
\newcommand{\xmark}{\ding{55}}%
\newcommand{\greentick}{✅}
\newcommand{\redcross}{❌}
\newcommand{\laugh}{😂}
\newcommand{\prayer}{🙏}

\begin{document}
\title{Auditing Apple’s DifferentialPrivacy.framework: Implementation Bugs, Misconfigurations, and Practical Risks}

\author{
    \IEEEauthorblockN{Rishav Chourasia\IEEEauthorrefmark{1}\IEEEauthorrefmark{2}, Ergute Bao\IEEEauthorrefmark{3}, Uzair Javaid\IEEEauthorrefmark{2}, Xiaokui Xiao\IEEEauthorrefmark{1}}
    \\
    \IEEEauthorblockA{
      \IEEEauthorrefmark{1}National University of Singapore \\ \IEEEauthorrefmark{2}Betterdata.ai \\\IEEEauthorrefmark{3}Mohamed bin Zayed University of Artificial Intelligence \\
      Email: \{rishav, uzair\}@betterdata.ai, baoergute8@gmail.com, xkxiao@nus.edu.sg
    }
}

\maketitle


\begin{abstract}

  Since 2016, Apple claims that the device analytics it collects to improve user experience are protected by differential privacy (DP). Running on over 2.35 billion Apple devices today,\footnote{Source: MacRumors report on \href{https://macrumors.com/2025/01/30/apple-active-devices-worldwide-record}{Apple's worldwide active device count}.} their DP framework gathers a variety of sensitive data ranging from contents of photos to COVID-19 vaccination status. However, despite strong community interest, Apple has not open-sourced its privatization algorithms, impeding independent researchers from verifying its privacy claims and testing for privacy violations. We perform \chgrep{the first}{a deep} audit of Apple's client-side DP framework \chgadd{on macOS Sonoma 14.2 and Sequoia 15.6} by reverse engineering its native binaries \chgdel{on macOS Sonoma and Sequoia} and \chgadd{building runtime interfaces to execute} Apple's deployed mechanisms. Our audit covers nearly all deployed algorithms, including Count Median Sketch (CMS) and Hadamard-CMS from Apple's white paper~\cite{apple2017learning}, and the Prio protocol for secure aggregation (SecAgg), which Apple uses for learning iconic scenes~\cite{apple_differential_privacy}.
  
  We report multiple issues. Every algorithm in Apple's framework that relies on floating-point noise fails to meet its advertised DP and zero-knowledge (ZK) proof guarantees. These violations happen because the DP framework uses insecure noise generators that are known to have floating-point vulnerabilities since 2012. We also discover that the SecAgg protocols in Apple's DP framework are configured with local DP disabled, uploading data without any local DP protection. Our audit provides evidence of DP violations in 5 out of 9 audited mechanisms, impacting 87\% of data collection in macOS Sonoma and 68\% in Sequoia. These vulnerabilities are exploitable by any party with access to pre‑aggregation logs. We found instances on the open internet of leaked iPhone logs that can be decoded to recover private information, such as domains visited in Safari and emojis typed on the iPhone keyboard. We responsibly disclosed these vulnerabilities to Apple; the concerned mechanisms are now deprecated.

\end{abstract}

\IEEEpeerreviewmaketitle


\section{Introduction}
\label{sec:intro}

Differentially private (DP) data collection has become a popular method for gathering user insights. Its applications range from enterprise use-cases like Google Trends~\cite{zhang2023differentially}, Meta's COVID-19 mobility reports~\cite{meta2020covid19}, and Windows OS's analytics~\cite{ding2017collecting}, to government tasks like the US Census Bureau's demographic data release~\cite{abowd2018us}. 
The privacy-enhancing computation market was valued at USD 5.2 billion in 2024.\footnote{Source: \href{https://market.us/report/privacy-enhancing-computation-market}{Privacy-enhancing computation market report}.}

When used correctly, DP algorithms for data-aggregation provide a mathematical guarantee that the contributing user's privacy remains protected. Such a guarantee is ensured by deliberately introducing carefully-calibrated noise locally in the device data before uploading it to the servers. Apple is arguably the industry leader in secure and privacy-preserving data aggregation technology. Having announced the adoption of DP as early as in 2016 for collecting device-analytics in iOS 10~\cite{applekeynote2016}, the company demonstrated its strong commitment with a white paper detailing its large-scale architecture for its DP framework~\cite{apple2017learning} in 2017. Since then, Apple has secured numerous patents on distributed learning and aggregate trend analysis with DP~\cite{bhowmick2018differential,bhowmick2019privatized,bhowmick2024private,bhowmick2023distributed,thakurta2017emoji,thakurta2017learning,friedman2020differential,sierra2019efficient}. Apple has also publicly emphasized its commitment to privacy through marketing and research communications~\cite{apple2025understanding,chadha2024differentially,feldman2021lossless,feldman2022private}. Today, Apple's DifferentialPrivacy.framework is present in every device running Apple's operating systems---iOS, iPadOS, macOS, watchOS, tvOS, and visionOS.\footnote{Sources: \href{https://theapplewiki.com/wiki/Filesystem:/System/Library/PrivateFrameworks\#:~:text=DifferentialPrivacy}{public wiki}, \href{https://www.developer.limneos.net/index.php?framework=DifferentialPrivacy.framework}{dyld header dumps}, \href{https://arxiv.org/abs/1709.02753}{prior research}, \href{https://www.apple.com/legal/privacy/data/en/device-analytics/}{device analytics} \& \href{https://www.apple.com/legal/privacy/data/en/intelligence-engine/}{Apple intelligence} privacy policies, \href{https://support.apple.com/en-sg/guide/tv/atvb66239fa1/tvos}{AppleTV user-guide} and more.}

Yet, unlike its competitors like Google, Apple has not open-sourced its DifferentialPrivacy.framework. Consequently, third party researchers have been prevented from validating the correctness of Apple's critical privacy algorithms, leading privacy experts to question the real-world effectiveness of the deployed techniques~\cite{Cyphers2017,HackerNews2016}. 
This lack of transparency hinders independent verification: as with cryptographic primitives, \emph{differentially private algorithms are difficult to implement correctly}---there are many examples of flawed implementations with subtle bugs that do not meet their configured DP guarantees~\cite{casacuberta2022widespread,mironov2012significance,lebeda2024avoiding,tramer2022debugging,lyu2016understanding}. 

This paper presents \chgrep{the first}{an} \emph{independent, deep-dive privacy audit of Apple’s DifferentialPrivacy.framework}, focusing on the versions deployed in macOS Sonoma 14.2 (\chgadd{released on Dec 11, 2023}) and Sequoia 15.6 (\chgadd{released on July 29, 2025}). Our work surgically deconstructs the black-box nature of the framework by decompiling its native binaries to extract Objective-C runtime headers. We then build custom executables that dynamically load the framework to execute Apple's native privatization algorithms, allowing us to perform a combination of static analysis and runtime observation. 
\chgadd{This goes beyond all prior analyses of Apple's DP algorithms or frameworks as by recovering executable interfaces to the shipped binaries themselves, we audit the native outputs on modern Apple OSs.}
Through this approach, we uncover \chgdel{implementation} defects that lead to practically exploitable vulnerabilities in several core privatization algorithms.
We not only design decoders that can de-noise and extract user information from analytics data destined for Apple’s servers, but we also demonstrate realistic exploits that external attackers can mount on leaked data. We found numerous real-world instances where these private device-analytics logs have leaked onto the open internet---through user forums, social media, and public code repositories
like GitHub---exposing private data through harmless-looking ciphertexts.


\emph{Our findings contradict Apple's claims regarding the privacy of collected device analytics through their DP framework}.
Using state-of-the-art (SOTA) privacy auditing techniques~\cite{ding2018detecting,zanella2023bayesian}, we provide \emph{statistically-significant evidence} that Apple's implementation of these critical mechanisms systematically violates their claimed differential privacy and zero-knowledge (ZK) proof guarantees. We find floating-point vulnerabilities in Laplace and Gaussian pseudorandom number generators (PRNGs) that are central to several DP and secure aggregation (SecAgg) algorithms actively used to collect device data for $187$ out of $276$ (i.e., $\approx 68\%$) property keys in the macOS Sequoia 15.6 and $160$ out of $185$ (i.e., $\approx 86\%$) keys in macOS Sonoma 14.2. 

More concerningly, we find that Apple’s SecAgg protocols are \emph{purposely configured to not be differentially private}. Although \emph{SecAgg is fully compatible with DP}~\cite{corrigan2017prio,rothblum2024pine} and the implemented protocols do support DP with an optional flag that can be enabled, our audit shows that the majority of configurations do not enable this protection. We demonstrate how any data collected this way can be \emph{reconstructed exactly on a malicious Apple server} (or by an external adversary, if leaked). We also highlight that the design of Apple's SecAgg protocols break a core tenet of multiparty computation (MPC) by sending all the cryptographic shares of users' data to a single (potentially untrusted) leader server.

\begin{figure}[t]
	\centering
  \makebox[\columnwidth][c]{%
	  \includegraphics[width=1.05\linewidth]{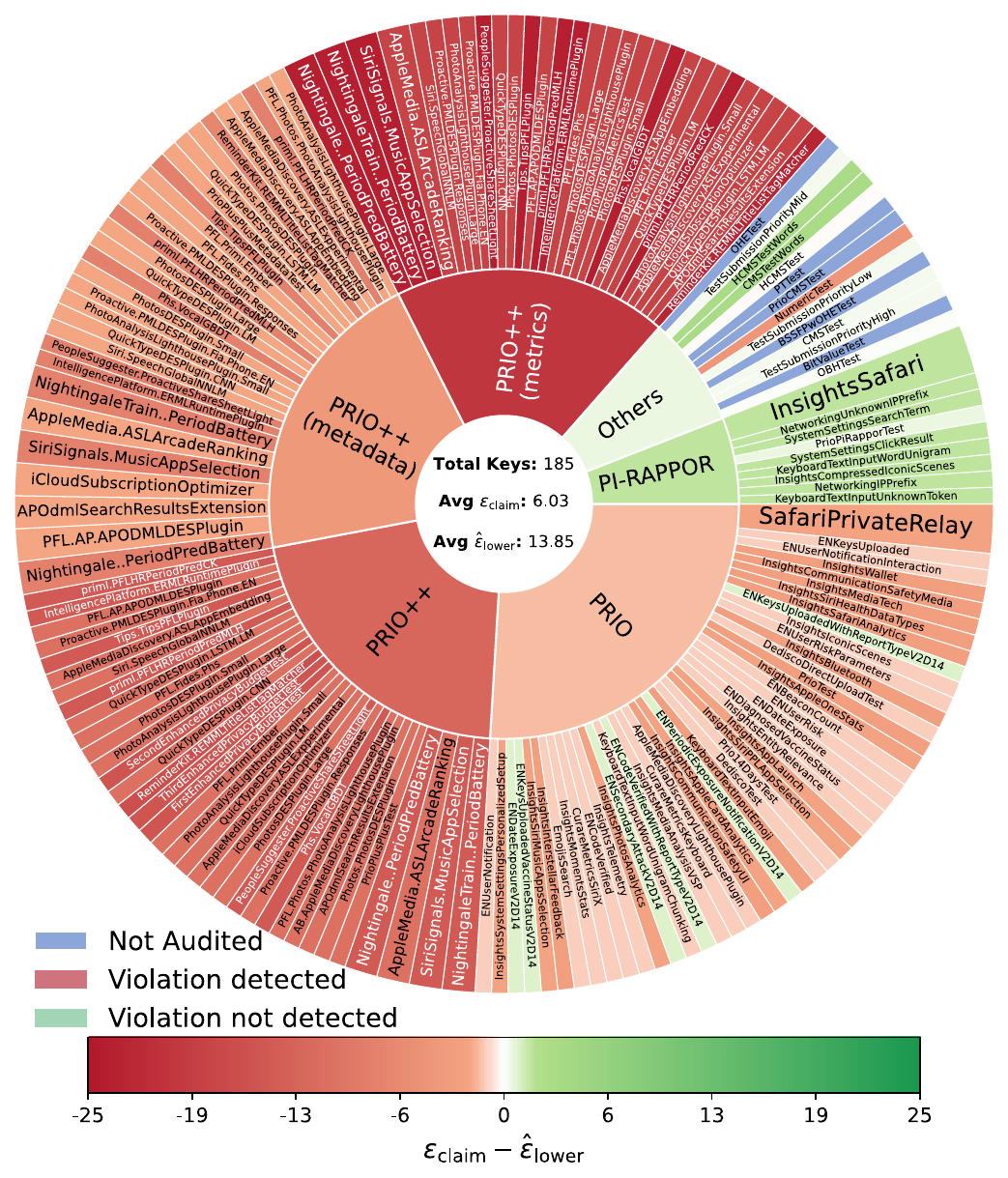}
  }
  \caption{Distribution of DP mechanisms identified within macOS Sonoma 14.2 (inner ring), paired with the specific property keys configured for data collection using these mechanisms (outer ring). The gradient represents the severity of the associated privacy violation. We quantify the severity as the difference between the claimed DP upper bound (as per the `\emph{PrivacyParameter}' attributes in the `\emph{com.apple.dprivacyd.keyproperties.plist}' configuration file) and the lower bound $\hat\varepsilon_\text{lower}$ on the \emph{true} differential privacy parameter for which we have 95\% confidence based on our auditing results.}
  \label{fig:audit_sunburst} 
\end{figure}

Finally, even for the few DP mechanisms that passed our audit, we find their configured DP budgets are excessively large, often exceeding $\eps=6$. \chgrep{Such high $\eps$ values significantly erode privacy; for instance, they permit membership inference attacks to succeed with a greater than $99.75\%$ chance.}{Such high $\eps$ values significantly erode privacy; in the standard binary hypothesis-testing interpretation of pure DP, an $\eps=6$ guarantee is weak enough to allow membership inference success as high as $e^\eps/(1+e^\eps) > 99.75\%$~\cite{kairouz2015composition}.}
Collectively, our findings reveal that Apple’s implementation of differential privacy does not fully match its public claims, raising questions about the company's commitment to the privacy of its users. \autoref{fig:audit_sunburst} shows the key-names of these collected data and the extent of the violations per our audit on macOS Sonoma 14.2. We detected most of these violations in Sequoia 15.6 as well.

\subsubsection*{Related Work}
Some of our findings on implementation errors are related to floating-point vulnerabilities in differential privacy. Mironov~\cite{mironov2012significance} first demonstrated that the inverse-sampling method for Laplace distributions introduces irregularities that break DP guarantees. Subsequent work~\cite{jin2022we,haney2022precision,lokna2023group} uncovered similar vulnerabilities in various sampling techniques for continuous distributions and developed privacy attacks. We specifically build upon the attacks proposed by Mironov~\cite{mironov2012significance} and Jin et al.~\cite{jin2022we}, extending their attack principles to target the Gaussian and Laplace distribution sampling methods used in Apple's binaries.

To empirically validate our findings, we employ \emph{differential privacy auditing} techniques, which estimate a mechanism's DP constants by conducting membership inference attacks~\cite{ding2018detecting,gilbert2018property,bichsel2021dp,steinke2024privacy,mahloujifar2024auditing}. We use the Bayesian auditing approach by Zanella-Béguelin et al.~\cite{zanella2023bayesian}, incorporating refinements from Nasr et al.~\cite{nasr2023tight} to derive mechanism-specific bounds on the true DP parameters of Apple's mechanisms.
\chgadd{Our contribution is to operationalize these auditing techniques for Apple's closed-source implementations by reverse engineering the framework, building executable interfaces, and designing mechanism-specific attacks and log decoders of real system outputs.}

\chgdel{Prior reverse-engineering attempt in 2017 by Tang et al.~\chgprotect{\cite{tang2017privacy}} via static analysis of Apple's earlier DP library in macOS 10.12 provided notable insights into Apple's system design and DP budget configuration/renewal policies.}
\chgadd{Prior analyses of Apple’s DP deployment address different mechanisms and questions. Gadotti et al.~\cite{gadotti2022pool} analyzed practical leakage in Apple’s Count Mean Sketch using pool-inference attacks derived from Apple’s published CMS design and parameters~\cite{apple2017learning}. Tang et al.~\cite{tang2017privacy} offered an important early study of Apple’s 2017 macOS 10.12 deployment, clarifying the framework’s system design and privacy-budget configuration and renewal policies. In contrast, we audit Apple’s current DifferentialPrivacy.framework at the implementation level: we reverse engineer the shipped binaries, construct executable interfaces to the deployed mechanisms, and evaluate their native outputs across a much broader and higher-budget deployment.}

\section{Preliminaries}
\label{sec:prelims}

Let $\M: \X \rightarrow \mathrm{Prob}(\Y)$ be a randomized algorithm that takes as input a record from the data universe $\X$ and produces a random outcome in the output space $\Y$.
\chgadd{In this paper, $\M$ is the device-side randomizer executed before any report is uploaded to Apple, i.e., the \emph{local} differential privacy (local DP) setting. By contrast, in \emph{central} (or \emph{global}) differential privacy, a trusted curator first collects raw data and only then releases a randomized aggregate or statistic~\cite{dwork2014algorithmic,bassily2015local}.}
Differential privacy ensures privacy in the form of \emph{plausible deniability}: an observer looking at an outcome $Y \sim \M(X)$ is unable to accurately determine whether the input was $X=\x$ or an alternate $X=\x'$.

\begin{definition}[Differential Privacy~\cite{dwork2006differential}]
  A mechanism $\M$ is $(\eps, \del)$-\emph{differentially private} ($(\eps, \del)$-DP) if for all $\x, \x' \in \X$, with the output distributions $\M(\x), \M(\x')$ denoted as $P, Q$ respectively, the following condition holds:
  \begin{equation}
    \forall S \subset \Y \ : \ P(S) \leq e^\eps \cdot Q(S) + \del.
  \end{equation}
\end{definition}
\chgadd{For the local-DP mechanisms studied here, this condition compares the distributions of the \emph{uploaded report} produced on a single user's device under two possible raw inputs $\x$ and $\x'$.}
For small $\epsilon$ and $\delta$, with $\delta < 10^{-5}$ and $\epsilon < 1$, an $(\epsilon,\delta)$-DP guarantee implies minimal privacy risk: changing an individual’s data changes the outcome probability by at most a factor of $e^1 \approx 2.718$ with over $99.999\%$ probability. 
Although each privatized sample carries limited information, their aggregation permits learning population statistics with meaningful accuracy. For example, estimating the population mean $\frac{1}{n} \sum_{i=1}^n X_i$ under local $(\eps, \del)$-DP for $X_i \in [0,1]$ yields an expected root-mean-square error of $O(\sqrt{\ln (1/\del)} \big/ \eps\sqrt{n})$ using the Gaussian mechanism~\cite{dwork2014algorithmic}. To contextualize this, even with $(\eps,\del) = (1, 10^{-5})$, aggregating data from one million users incurs less than $0.373\%$ error, despite each user's submission having an expected root-mean-square error of over $373\%$.

\subsubsection*{Secure Aggregation}
The objective of \emph{secure aggregation} is to compute the global sum $A = \sum_{i=1}^n \M(X_i)$ without revealing any individual $\M(X_i)$ any more than what could be inferred from the aggregate $A$. Achieving this requires extending the trust boundary beyond the user's device to a group of $m$ external servers, each assumed to be \emph{honest but curious}. In this setting, every user splits their value $\M(X_i)$ into $m$ cryptographic fragments $[\M(X_i)]_1, \cdots, [\M(X_i)]_m$, sending one fragment to each server. Each server locally aggregates its received secret-shares as $[A]_j = \sum_{i=1}^n [\M(X_i)]_j$. When the servers jointly combine their partial aggregates, the true sum is recovered as $\sum_{j=1}^m [A]_j = A$, without exposing any individual $\M(X_i)$.

Vanilla SecAgg protocols are susceptible to poisoning attacks where one user can corrupt the aggregate $A$ by sending malicious shares to the servers. Verifiable SecAgg protocols like Prio~\cite{corrigan2017prio} provide robustness against such attacks by requiring users to additionally send a \emph{secret-shared non-interactive proof} (SNIP), which is a zero-knowledge (ZK) proof of correctness of their submitted data. The servers can use these secret-shares of a SNIP to collaboratively check the syntactic validity of a user's submission, without ever seeing their data in the clear.

SecAgg protocols exhibit a strong synergy with local DP mechanisms: the aggregate $A$ enjoys a considerably stronger amplified DP guarantee than individual $\M(X_i)$, \emph{provided that we trust that at least one server remains honest}~\cite{mcmillan2022private}.

\subsubsection*{Privacy Auditing}
While DP provides a theoretical upper bound on information leakage---typically derived through formal mathematical analysis---these guarantees may fail in practice if the analysis is flawed or the implementation is buggy. Such discrepancies are not rare, and numerous works have been devoted solely to uncovering analytical or implementation errors in published mechanisms~\cite{chen2015privacy,lyu2016understanding,mironov2012significance,casacuberta2022widespread,lebeda2024avoiding}. To complement theoretical proofs, \emph{privacy auditing} techniques have emerged as a practical means to obtain empirical lower bounds on the true DP parameters. By running privacy attacks on the implementation directly, these audits can reveal when empirical lower bounds contradict theoretical upper bounds, or otherwise assess the tightness of existing analyses~\cite{lokna2023group,steinke2024privacy,ding2018detecting,bichsel2021dp,mahloujifar2024auditing,zanella2023bayesian}. We describe the methodology used in this paper in {\def\sectionautorefname{Appendix}\autoref{sec:app_audit}}.


\section{System Overview of Apple's DP Framework}
\label{ssec:design}

This section describes Apple's client-side system architecture for collecting device analytics data using differential privacy. Our analysis is based on an inspection of system files, configurations, binaries, and logs from macOS Sonoma 14.2 (build 23C64) and Sequoia 15.6 (build 24G84). 

\subsubsection*{Framework's Workflow}
At the heart of the architecture is DifferentialPrivacy.framework, found within `\emph{/System/Library/PrivateFrameworks}', which is a bundle containing a \emph{dynamic library} (DYLIB) and its associated assets. This library provides the full suite of DP and SecAgg functionalities, accessible through its \hdr{_DPServer.h}{DPServer} class, which runs as a system daemon named `\emph{dprivacyd}'.

Upon startup, the `\emph{dprivacyd}' system daemon initializes several periodically scheduled background processes. These processes are essential for tasks like managing the privacy budget, maintaining and cleaning up local storage, and generating report logs. Once initialized, the server begins listening for XPC (Cross-Process Communication) requests from other applications that want to record analytics data. It can record a variety of analytics data-types, including strings, boolean values, and numerical data. To ensure data integrity and security, the XPC listener only accepts connections from applications holding a special entitlement reserved exclusively for Apple's own applications.

\begin{figure}[t]
	\centering
  \includegraphics[width=\linewidth]{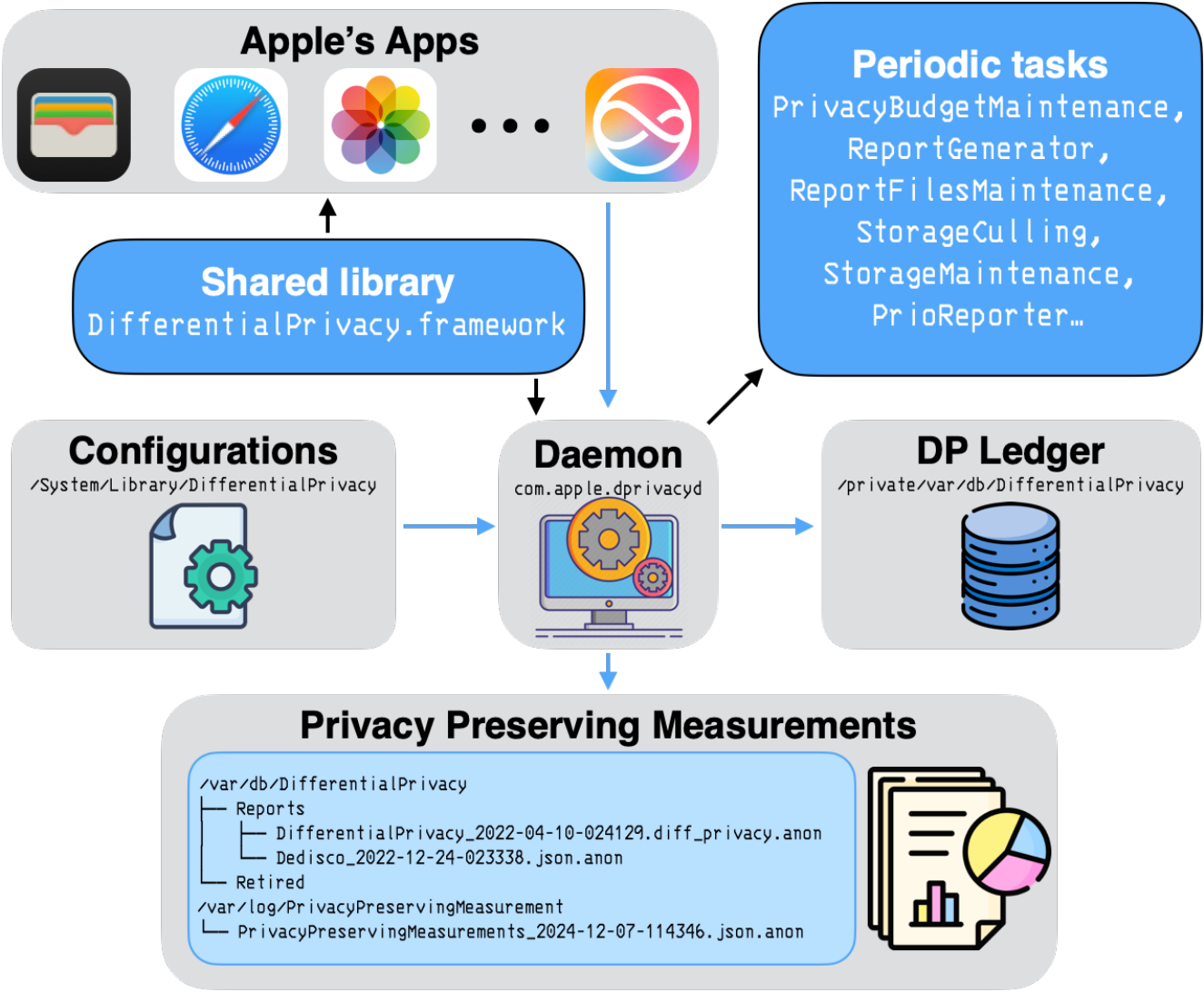}
  \caption{Overview of the design of DifferentialPrivacy.framework.}
  \label{fig:system_overview} 
\end{figure}

The `\emph{dprivacyd}' daemon relies on a set of hardcoded file paths---stored within the DYLIB's binary in the \hdr{_DPStrings.h}{DPStrings} class---to locate key resources like the database maintaining the DP ledger, the directory where reports are generated, and configurations for privacy budget and algorithm parameters. This configuration data governs the privatization process: when an app requests to record data, `\emph{dprivacyd}' uses these internal configs to determine the specific privatization algorithm, its corresponding parameters, the amount to debit in the DP ledger, whether the analytics report should be queued locally or uploaded immediately, and other settings. \autoref{fig:system_overview} provides an overview of how the components of this DP framework interact to collect device analytics data.

\subsubsection*{Privatization Algorithms}
DifferentialPrivacy.framework also contains the core implementations of Apple's Local DP mechanisms and secure aggregation protocols, including \hdr{_DPCMSRandomizer.h}{Count Median Sketch}~\cite{apple2017learning}, \hdr{_DPPiRapporAlgorithm.h}{PI-RAPPOR}~\cite{feldman2021lossless}, and \hdr{_DPPrioAlgorithm.h}{Prio}~\cite{corrigan2017prio}. The `\emph{dprivacyd}' service uses these algorithms to privatize the sensitive analytics data requested for upload by authorized apps. \chgrep{Our analysis found 14 such algorithms in production on macOS Sonoma 14.2, where our DYLIB hijack (described next) successfully interfaced with nine (9) of them. Following the upgrade to macOS Sequoia 15.6, some of these algorithms were replaced, with our DYLIB hijack working for seven (7) out of the 14 algorithms.}{Our analysis found 14 such algorithm families in production on macOS Sonoma 14.2. Our DYLIB hijack (described next) successfully interfaced with the nine families that drive almost all active collections studied in this paper: NumberRandomizer, CMS, HCMS, PI-RAPPOR, OneBitHistogram, Prio, Prio++, Prio++Metadata, and Prio++Metrics. The remaining Sonoma families were wrapper or specialized variants, including PrefixTree, BSSFPwOHE, and PrioCMS, that we did not observe driving any active user-data collection in the analyzed deployment (cf.~blue wedges in~\autoref{fig:audit_sunburst}'s outer circle). 

Following the upgrade to macOS Sequoia 15.6, some historical mechanisms were replaced by newer algorithms like Prio3 and PINE that according to Apple follow public designs~\cite{rothblum2024pine,irtf-cfrg-vdaf-17,ietf-ppm-dap-16} (cf.~{\def\subsectionautorefname{Appendix}\autoref{ssec:apple_deployment_status}}); our DYLIB hijack worked for seven (7) of the 14 families there. Although we did not instrument these two new client-side variants, the same dynamic-loading methodology should extend to them.}

\subsubsection*{DP Budget Ledger}
All recorded values, the corresponding algorithm parameters, and timestamps are locally stored in an SQLite database named `\emph{DifferentialPrivacyClassC.db}', found in `\emph{/private/var/db/DifferentialPrivacy/}'. This database also includes a table named `\emph{ZPRIVACYBUDGETRECORD}', which tracks the DP budget balance for every configured property collected across the system. 

\subsubsection*{Analytics Logs}
The database directory also contains `\emph{Reports/}' and `\emph{Retired/}' subdirectories where the analytics logs are generated. Created by the \hdr{_DPReportGenerator.h}{report-generator} periodic task, these logs are plain text files with prefixes like `\emph{Dedisco}', `\emph{DifferentialPrivacy}', or `\emph{PrivacyPreservingMeasurements}'. In iOS and iPadOS, these staged reports can be viewed by navigating to Settings $\rangle$ Privacy \& Security $\rangle$ Analytics \& Improvements $\rangle$ Analytics Data. A subsequent \hdr{_DPReportFilesMaintainer.h}{report-maintainer} task periodically rotates these files into the `\emph{Retired/}' subdirectory before deleting older records.

\subsubsection*{System Integrity Protection}
We note that Apple enabled \emph{System Integrity Protection} (SIP) on many associated directories for its DP framework following the reverse-engineering work by Tang et al.~\cite{tang2017privacy}. In order to examine those paths, we had to disable SIP on our testing devices.

\section{Hijacking DifferentialPrivacy.framework}
\label{ssec:linking-appledp}

\begin{figure*}[t]
\begin{minipage}[t]{\linewidth}
\begin{lstlisting}
#import <Foundation/Foundation.h>
#import "_DPNumberRandomizer.h"   // Header extracted from `dyld_shared_cache' using class-dump

int main() {
    /* Dynamically load the DP framework */
    NSString *path = @"/System/Library/PrivateFrameworks/DifferentialPrivacy.framework";
    NSBundle *framework = [[NSBundle alloc] initWithPath:path];
    [framework load];

    /* Initialize and run DifferentialPrivacy.framework's NumberRandomizer */
    double input = 0, epsilon = 1, range = 1;
    id mech = [[NSClassFromString(@"_DPNumberRandomizer") alloc] initWithRange:range epsilon:epsilon];
    NSLog(@"Sample: %@", [mech randomize:@(input)]);
} // > gcc -framework Foundation laplace.m -o laplace
\end{lstlisting}
\end{minipage}
\caption{\label{fig:number_randomizer_code}A program demonstrating how to dynamically load Apple's DifferentialPrivacy.framework and access its \hdr{_DPNumberRandomizer.h}{NumberRandomizer} functionality.}
\end{figure*}

Apple's operating systems are built on a foundation of both \emph{public and private frameworks}. Public frameworks, such as MediaPlayer.framework and UIKit.framework, are documented and supported for all third-party developers. Conversely, private frameworks like DifferentialPrivacy.framework are internal components reserved exclusively for Apple's own applications and system services. Apple actively prevents third-party use of these internal components through several mechanisms: refusing to release documentation, scanning App Store submissions for linked private frameworks, and requiring special entitlements within the application sandbox to access its resources.

To use a framework, an executable must be compiled against its header files (with a .h extension), which declare the available classes and functions within its DYLIB. Programs can incorporate these functionalities in two ways. The first is \emph{dynamic linking}, where the static linker includes the library's symbols as dependent libraries during compile time. When the program launches, the \emph{dynamic link editor} (or \emph{dyld} for short) handles the automatic loading and symbol resolution of these shared libraries into memory. The second and more flexible approach is \emph{dynamic loading}, where the static linker adds no dependency information. Instead, the program code itself explicitly loads the library at runtime (e.g., by calling the \emph{load()} method) to resolve the necessary symbols. The \emph{dyld} handles both scenarios, ensuring dependent libraries are loaded at launch and managing runtime requests for dynamically loaded libraries. Crucially, as long as the framework is properly linked or loaded, even a third-party program can access and use the functionalities implemented in Apple's private frameworks.

Apple provides neither the documentation nor the header files for its DifferentialPrivacy.framework. However, using open-source tools such as class-dump and ipsw, we extract the header files directly from the framework's binary.\footnote{All frameworks' binaries used to be located inside respective subdirectories in `\emph{/System/Library/}'. However, since macOS Ventura 13, all system binaries are consolidated into a single `\emph{dyld\_shared\_cache}' file. We extract the DP framework's binary from it using dyld-shared-cache-extractor.} \chgadd{Recovered headers are available in the associated artifact~\cite{apple_dp_artifact}. }We rely on \emph{dynamic loading} in our programs to interface with DifferentialPrivacy.framework to avoid the complaints by the static linker regarding unresolved external references during compilation. This flexibility comes with a price as errors due to incorrect declarations in the headers or missing symbols may result in a runtime crash. We debug such crashes by inspecting the relevant parts of the DYLIB's binary with decompiling software like Hopper, \chgdel{IDA-pro}\chgadd{IDA Pro}, and Binary Ninja. \autoref{fig:number_randomizer_code} illustrates an example program that successfully executes the \hdr{_DPNumberRandomizer.h}{NumberRandomizer} which implements the Laplace mechanism. With some hit-and-trial, we built an interface to run almost all privatization mechanisms in DifferentialPrivacy.framework. \chgadd{We validated recovered declarations against both binary evidence (e.g., constants, types/casts, selector usage, and branch structure) and runtime behavior through executable instrumentation; mismatches typically caused runtime failures that we iteratively resolved. This lets us instrument Apple's shipped implementations directly, enabling tight privacy audits and practical reconstruction attacks from logs, as shown later.}

\subsubsection*{LLM's Aid in Binary Analysis}
Analyzing decompiled code can be challenging due to many reasons, like changes in logic due to the optimizations made by the compiler and the absence of meaningful function/variable names. To understand some of the complex parts of the decompiled binary, we sought assistance from Large Language Models (LLMs), specifically ChatGPT 4o and Gemini 2.5 Pro models, in deciphering the underlying code. \chgadd{Concretely, we used LLMs to help interpret decompiled Objective-C code: we pasted a snippet from Hopper, IDA Pro, or Binary Ninja together with our suspected functionality and asked for a cleaner Objective-C or pseudocode rewrite. We treated these outputs only as hypotheses. Any interpretation we relied on had to match binary evidence such as constants, types/casts, and branch logic, and also be consistent with runtime behavior observed through our executable instrumentation. When an LLM interpretation disagreed with this evidence, we discarded it.} \chgrep{By providing these models with additional contexts regarding the suspected functionality, we found their insights to be helpful in our analysis.}{When provided with accurate context regarding a snippet's suspected functionality, we found these LLMs' interpretations to be helpful in our analysis.}

\subsubsection*{Circumventing Apple's Security}
Our approach bypasses the security boundary created by Apple's \emph{Sandbox} and \emph{Hardened Runtime} features that protect the `\emph{dprivacyd}' daemon. This is because the framework logic resides within DifferentialPrivacy.framework's binary, while `\emph{dprivacyd}' is merely a service intended to act as its gatekeeper via the XPC protocol. By employing dynamic loading to bring the framework's code directly into our application's process space, we eliminate the need to establish a secure XPC connection with `\emph{dprivacyd}'. Instead, our program directly calls the privacy functions implemented within the framework's DYLIB, avoiding all the high-level security checks, entitlements, and system service restrictions that would otherwise be enforced against any third-party app attempting to communicate with the protected daemon. 


\section{Threat Models and Attack Scenarios}
\label{ssec:threat}

Our analysis considers multiple threat models from the perspectives of both Apple’s servers (assuming them to be dishonest, collusive, and curious) and external adversaries.

\subsubsection*{Reconstruction Attacks}
In our primary threat model, we assume that the adversary can access the privatized analytics records generated on an Apple device. The adversary's objective is to \emph{decode} meaningful personal information from these analytics records. The attacks are conducted in an \emph{offline setting}---without interacting with the users. To extract personal data, the adversary exploits vulnerabilities in the privatization algorithms used to generate the records. This threat model applies in the following scenarios:

\begin{enumerate}[label=\textbf{\Alph*})]
  \item \textbf{Apple's servers.} In its ``\href{https://www.apple.com/in/privacy/approach-to-privacy/index.html}{Approach to Privacy}'' page, Apple claims that personally identifiable information (PII) is ``removed from reports before they're sent to Apple or protected by [...] Differential Privacy.'' We examine this claim by testing whether the uploaded records indeed provide sufficient privacy such that even Apple cannot infer personal user information from the collected data.

  \item \textbf{Third-party attackers.} This scenario involves an external adversary who gains unauthorized access to analytics logs originating from iPhones, iPads or other Apple devices. \chgadd{This is a post-export threat model: we do not assume that arbitrary third-party apps can bypass Apple's sandbox and read these files directly on-device. Rather, the adversary gains access after user-exportable diagnostics or staged analytics logs leave the device, for example through log-stealer malwares, diagnostics sharing or public leaks. Once this happens, the attack is fully offline.}
  In particular, we observed over fifty publicly-posted instances of such logs on Apple's official forum, Pastebin, GitHub, Reddit, and even on Instagram. (To avoid resurfacing potentially sensitive user data, we omit direct URLs.)
  \chgadd{Unlike screenshots or photos, these logs contain machine-readable encoded measurements whose sensitivity is not obvious to users.} The ability to decode such logs presents a serious privacy risk to unsuspecting users and is difficult for Apple to mitigate, as these logs are already public.

\end{enumerate}

\subsubsection*{Membership-Inference Attacks}
In this threat model, the attacker knows that the mechanism's input is either $x_0$ or $x_1$ and seeks to determine which of the two was used based on the observed output. By repeating this test across many independent runs, where the input is randomly chosen between $x_0$ and $x_1$, we measure the attacker's prediction accuracy to derive empirical lower-bound estimates on the the true level of DP of the mechanism. We highlight that this threat model is not intended to reflect a real-world attack setting; rather, it provides a controlled experiment for auditing the DP guarantees claimed by Apple’s framework.


\section{Floating-Point Vulnerabilities}
\label{ssec:fp_vuln}

Many privatization algorithms in Apple's DP framework rely on continuous distributions as their core components. However, the \emph{finite-precision} of digital systems prevents exact representation of continuous random variables. 
Standard numeric libraries like random.h in C++ and NumPy or PyTorch in Python approximate these variables using IEEE 754 floating-point arithmetic, which introduces subtle irregularities. When used to generate noise in DP applications, these imperfections can break the privacy guarantees~\cite{mironov2012significance}. 
As such, modern DP frameworks employ specialized, cryptographically secure noise generation methods~\cite{google_secure_noise,holohan2024securing,keller2024secure}.

\subsection{Laplace Mechanism} 
\label{ssec:laplace}

We find that the \hdr{_DPLaplaceNoiseGenerator.h}{LaplaceNoiseGenerator} in Apple’s DP framework, used by their \hdr{_DPNumberRandomizer.h}{NumberRandomizer} class implementing the Laplace mechanism~\cite{dwork2006calibrating}, employs the inverse transform sampling method~\cite{gentle2003random} as outlined in \autoref{alg:lap}. This method leverages the property that any random variable $Y$ with cumulative distribution function (CDF) $F(t) = \Pr[Y \leq t]$ can be generated from a uniformly distributed random variable $U \sim [0, 1]$ through the transformation $ Y \leftarrow F^{-1}(U)$. This follows from the fact that for all $t \in \R$, 
\begin{equation}
  \Pr[F^{-1}(U) \leq t] = \Pr[U \leq F(t)] = F(t).
\end{equation}
\begin{algorithm}[!htb]
  \DontPrintSemicolon
  \caption{inverse\_sampler for $\lap(\lambda)$}\label{alg:lap}
  \KwData{Scale parameter $\lambda$}
  \KwResult{A single sample $Y \sim \lap(\lambda)$}
   uniformly sample $U \sim [0,1]$\;
   let $F^{-1}(u) \eqdef \mathrm{sign}\left(\frac{1}{2} - u\right) \cdot \lambda \ln\left(1 - 2 \left\vert u - \frac{1}{2}\right\vert\right)$\;
   $Y \leftarrow F^{-1}(U)$ \pccomment{$F^{-1}(\cdot)$ is the inverse CDF of $\lap(\lambda)$} \;
   \textbf{Return:} $Y$\;
\end{algorithm}

\subsubsection*{Vulnerability}
The framework implements the arithmetic in \autoref{alg:lap} using double-precision numbers. By the IEEE standard, this \textit{double} datatype allots 1 bit for sign, 52 bits for the mantissa, and 11 bits for the exponent, and together they can represent numbers of the following form.
\begin{align*}
  Y \text{ in memory } &: \underbracket{s}_{\text{sign}} \ \underbracket{eeeeeeeeeee}_{\text{exponent}\ E} \ \underbracket{b_{51} b_{50} ...b_2 b_1 b_0}_{\text{mantissa}} \\
  \text{Real value of }Y  &: (-1)^s \left(1 + \sum_{i=1}^{52}b_{52 - i}\cdot2^{-i}\right) \times 2^{E-1023} 
\end{align*}
To produce a sample $Y \sim \lap(\lambda)$, the pseudorandom generator first samples a random 32-bit unsigned integer and divides it by its max value $2^{32} - 1$ to get a double $U$ in range $[0,1]$. Then, it computes its inverse CDF value $F^{-1}(U)$ using floating-point arithmetic and returns it. 

Back in 2012, Mironov~\cite{mironov2012significance} noted that the double $U$ produced this way does not have support over all the 64-bit doubles in the range $[0,1]$. Furthemore, even if care was taken to include all the doubles in the support of $U$, Mironov~\cite{mironov2012significance} noted that applying the inverse CDF $F^{-1}(\cdot)$ in floating-point arithmetic still results in plenty of double values missing from the support of $Y$. As a result, when used in a Laplace mechanism $\M_\lap(x) = q(x) + \lap(\Delta_q / \eps)$ to privatize a query $q: \mathcal{X} \rightarrow [0,\Delta_q]$, this generator leads to  an $\eps$-DP violation: because certain double values can never occur for one input but may appear for another,  detecting such an \emph{infeasible} output provides \emph{incontrovertible} evidence of which input was used (say, among $x_0$ or $x_1$).

\subsubsection*{Membership-Inference Attack}
Based on the discovered floating-point vulnerability, we design an efficient attack against \chgdel{Apple's} $\M_\lap$ \chgadd{as implemented by Apple's} \chgdel{(in }NumberRandomizer\chgdel{)}. Described in \autoref{alg:lap_att}, our attack tests for the infeasibility of a candidate input $\mu$ given an output $Y$ of the mechanism.

\begin{algorithm}[!htb]
  \DontPrintSemicolon
  \caption{$\phi_\lap(Y; \mu, \lambda)$: Is $\mu$ infeasible?}
  \label{alg:lap_att}
  \KwData{Candidate mean $\mu$, output $Y$, scale $\lambda$}
  \KwResult{\textbf{True} or \textbf{False}}
   let $F(y) \eqdef \frac{1}{2} \left(1 + \mathrm{sign}(y) \cdot \left(1 - e^{-\frac{\left\vert y\right\vert}{\lambda}}\right)\right)$ \;
   let $F^{-1}(u) \eqdef \mathrm{sign}\left(\frac{1}{2} - u\right) \cdot \lambda \ln\left(1 - 2 \left\vert u - \frac{1}{2}\right\vert\right)$\;
   \textbf{Return:} $\mu + F^{-1}(F(Y - \mu)) \neq Y$\;
\end{algorithm}

Mathematically, this algorithm should always return \textbf{False} regardless of the choice of $\mu$ or $Y$, because $F^{-1}(F(\cdot))$ is an identity function. However, floating-point arithmetic changes this. If output $Y$ was not sampled from $\mu + \lap(\lambda)$, then there is a considerable chance ($\approx 78\%$) that the condition in line 3 evaluates to \textbf{True}. This happens because the output $Y$ might not be in the distribution of $\mu + \lap(\lambda)$. That is, there might not be any double $U \in [0,1]$ such that $\mu + F^{-1}(U)$ precisely equals $Y$. To check for this, \autoref{alg:lap_att} streamlines the search by only testing $U = F(Y - \mu)$, which yields a value closest to $Y$. Conversely, if $Y$ was in fact sampled from $\mu + \lap(\lambda)$, the test always finds a match, i.e., $\phi_\lap(\mu + \lap(\lambda); \mu, \lambda) = \textbf{False}$. 

\subsubsection*{Reconstruction Attack}
The accuracy of our test $\phi_\lap$ is $88.2\pm1.02\%$ regardless of the choice of DP parameter $\eps$, which is more than enough for a practical attack. But we can also boost the accuracy to $\approx100\%$ by drawing multiple samples $Y_1, \cdots, Y_m$ from the Laplace mechanism and testing whether any one of them is infeasible for a given input, i.e. $\bigvee_{i=1}^m \phi_\lap(Y_i;\mu, \lambda)$. At this accuracy level, we can effectively reconstruct the input perfectly from the noisy outputs. We find that with access to as few as $m=5$ samples from NumberRandomizer with a DP guarantee of $\eps=0.2$ (i.e., $1$-DP guarantee on composition) allows us to accurately infer the collected property ``\emph{com.apple.datatype.age}'' in the range $\{0,1, \cdots, 100\}$ with $90.4 \pm 0.91\%$ probability.

\subsubsection*{Privacy Auditing Result}
\begin{figure}[t]
	\centering
  \includegraphics[width=\linewidth]{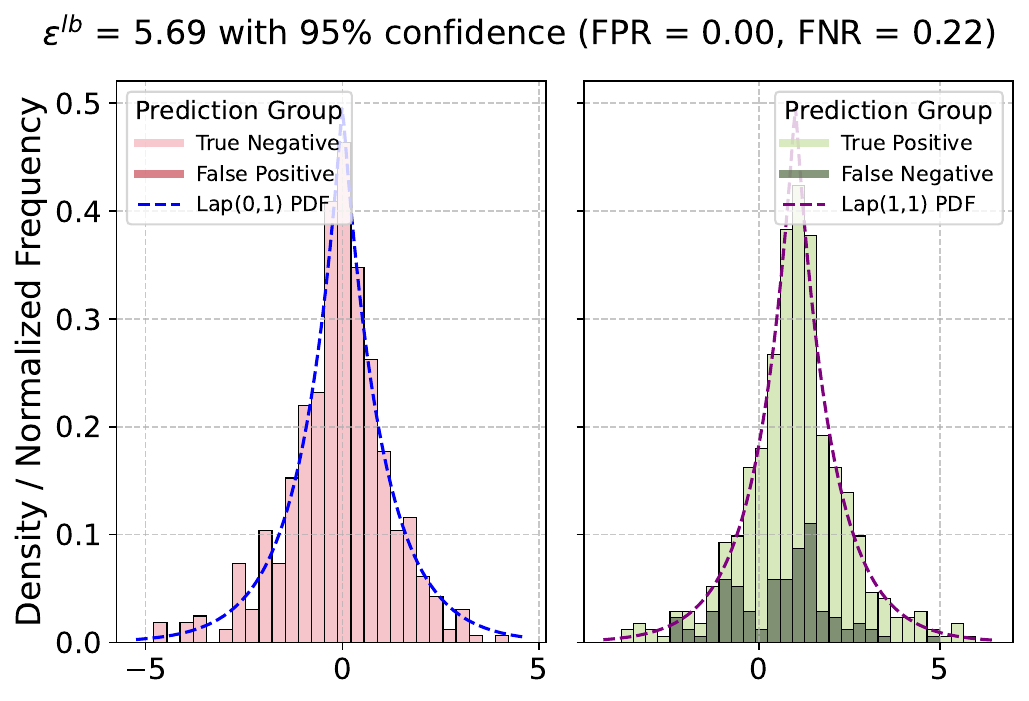}
  \caption{Result of our audit on Apple's NumberRandomizer with DP parameter set to $\eps=1$ detects a DP violation. We run the mechanism $n=1000$ times and use test $\phi_\lap(y;\mu=0, \lambda=1)$ to predict whether the input was $x_0=0$ or $x_1=1$. The audit used the family $\Fclass_\lap$ of Laplace mechanisms' trade-offs (cf. \autoref{rem:gauss_lap_family}) and the significance level was set to $\gamma = 0.05$.}
  \label{fig:laplace_result} 
\end{figure}

On running our DP audit (cf. \autoref{alg:zb_auditor} in {\def\sectionautorefname{Appendix}\autoref{sec:app_audit}}), we find that NumberRandomizer violates its DP guarantee. In \autoref{fig:laplace_result}, we show that our membership test $\phi_\lap$ achieves a false-negative rate (FNR) of $22.42 \pm 0.38\%$ with a false-postive rate (FPR) of $0.00 \pm 0.02 \%$, which constitutes statistically significant evidence that the true DP parameter exceeds
$\eps^\mathrm{lb} = 5.69$. Given that we configured the mechanism's parameters to $\eps=1$ and sensitivity $\Delta = 1$, our audit confirms a DP violation stemming from floating-point vulnerabilities.

\subsection{Gaussian Mechanism}
\label{sssec:prio}

One of Apple’s most widely used PRNGs in the framework, \hdr{_DPGaussianPRNG.h}{GaussianPRNG}, also suffers from floating-point vulnerabilities. It implements the \emph{Marsaglia polar} method to generate pairs of independent Gaussian samples as shown in \autoref{alg:gauss}. This method draws independent uniform variable pairs $(U_1, U_2) \in [-1, 1]$ from integer samples $V_1, V_2$, until $R = U_1^2 + U_2^2 \leq 1$. Since points $(U_1, U_2)$ are uniformly distributed within the unit circle, the normalized vector $(\frac{U_1}{\sqrt{R}}, \frac{U_2}{\sqrt{R}})$ is uniformly distributed on the unit circle, while the radius term $\rho = \sqrt{-2\log(R)}$ provides the correct radial scaling for the bivariate normal distribution. Thus, multiplying the two gives 
  $(Z_1, Z_2) = \left(\frac{U_1}{\sqrt{R}} \rho, \frac{U_2}{\sqrt{R}} \rho\right)$, 
whose coordinates are independent standard normal variables.

\begin{algorithm}[!htb]
  \DontPrintSemicolon
  \caption{marsaglia\_sampler for $\mathcal{N}(\mu,\sigma^2)$}\label{alg:gauss}
  \KwData{Random integers $V_1,V_2 \in [-2^{31}, 2^{31} - 1]$ s.t. $V_1^2 + V_2^2 \leq 2^{62}-1$, mean $\mu$, variance $\sigma^2$}
  \KwResult{Two samples $Y_1, Y_2 \sim \mathcal{N}(\mu, \sigma^2)$}
   $R \leftarrow (V_1^2 + V_2^2) / (2^{62} - 1)$\;
   $U_1 \leftarrow V_1/(2^{31}-1)$ and $U_2 \leftarrow V_2/(2^{31} - 1)$\;
   $Z_1 \leftarrow \frac{U_1}{\sqrt{R}} \sqrt{-2 \log(R)}$ and $Z_2 \leftarrow \frac{U_2}{\sqrt{R}} \sqrt{-2 \log(R)}$\;
   \textbf{Return:} $\mu + \sigma Z_1, \mu + \sigma Z_2$\;
\end{algorithm}

\subsubsection*{Vulnerability}
Similar to the Laplace noise generator, Apple's DP framework implements GaussianPRNG using floating-point arithmetic, which we discover to be vulnerable as well. Upon examining mechanisms like Prio++ and PINE, which rely on GaussianPRNG to privatize numeric vectors, we observe that both generated samples are always used sequentially. We exploit this in~\autoref{alg:gauss_att} by designing an infeasibility test that determines whether a candidate mean $\mu$ is feasible for an output pair $(Y_1, Y_2)$.
\begin{algorithm}[!htb]
  \DontPrintSemicolon
  \caption{$\!\phi_\gaus(Y_1, Y_2;\! \mu,\! \sigma^2)$: Is $\mu$ infeasible?}
  \label{alg:gauss_att}
  \KwData{Candidate mean $\mu$, output pair $(Y_1,Y_2)$, variance $\sigma^2$, testing window $k$}
  \KwResult{\textbf{True} or \textbf{False}}
   $Z_1 \leftarrow (Y_1 - \mu) / \sigma$ and $Z_2 \leftarrow (Y_2 - \mu) / \sigma$\;
   $R \leftarrow \exp\left(- \frac{Z_1^2 + Z_2^2}{2} \times \frac{(2^{31}-1)^2}{2^{62}-1}\right)$\;
   $V_2 \leftarrow \mathrm{sign}(Z_2)\cdot \mathrm{int}\left(\sqrt{\frac{R\cdot (2^{62} - 1)}{\vert Z_1 / Z_2 \vert^2 + 1}} \right)$\;
   $V_1 \leftarrow \mathrm{sign}(Z_1) \cdot \mathrm{int}\left(V_2 \left\vert\frac{Z_1}{Z_2}\right\vert\right)$\;
   \For{$V_1' \in \{V_1 - k, \cdots, V_1 + k\}$}{
     \For{$V_2' \in \{V_2 - k, \cdots, V_2 + k\}$}{
       $Y_1', Y_2' \leftarrow \mathrm{marsaglia\_sampler}(V_1', V_2', \mu, \sigma)$\;
       \If{$Y_1' \neq Y_1$ or $Y_2' \neq Y_2$}{
         \textbf{Return:} \textbf{True}\;
       }
     }
   }
   \textbf{Return:} \textbf{False}\;
\end{algorithm}

\subsubsection*{Membership-Inference Attack}
The above test works like the Laplace membership test: given a candidate mean $\mu$ we try to recover the integer pair $(V_1, V_2)$ that would have yielded the observed output $(Y_1, Y_2)$, then check whether the polar sampler with those integers yields the same outputs. Following the attack of Jin et al.~\cite{jin2022we}, we recover $R$ as:
\begin{align*}
  &- \frac{Z_1^2 + Z_2^2}{2\log(R)} =  \frac{V_1^2 + V_2^2}{R(2^{31}-1)^2} = \frac{2^{62}-1}{(2^{31}-1)^2} \\
  \implies & R = \exp\left(- \frac{Z_1^2 + Z_2^2}{2} \times \frac{(2^{31}-1)^2}{2^{62}-1}\right)
\end{align*}

In principle, testing whether $R \cdot (2^{62}-1) = V_1^2 + V_2^2$ is integer would suffice (the basis of Jin et al.~\cite{jin2022we}'s attack). However, Apple's implementation casts the 64-bit doubles $Y_1, Y_2$ to 32-bit floats before returning, losing $\approx 28$ bits of precision. This makes attacks much harder: the reconstructed $R$ effectively always integral and breaks that simple test.

To overcome this, we augment the attack. We also use the ratio $Z_1 / Z_2 = V_1 / V_2$ alongside the reconstructed $R$ to try and reconstruct $(V_1, V_2)$ as closely as possible. Unfortunately because of the precision loss, many nearby integer pairs can map to the same observed $(Y_1, Y_2)$. Our attack focuses on the \emph{size of this surrounding area}---if the candidate $\mu$ is correct, this ambiguity region should be large. Concretely, we test a window $V_1'\in \{V_1-k, \cdots V_1 +k\}$, $V_2' \in \{V_2 - k, \cdots, V_2 + k\}$ and verify whether the Marsaglia sampler with each $(V_1', V_2')$ reproduces $(Y_1, Y_2)$. If any neighbour produces a different output, we mark $\mu$ as infeasible. With a window size $k=80$ this test achieves an FPR of $0.00\pm0.00\%$ and a TPR of $0.25 \pm 0.03\%$, and an overall balanced accuracy of $50.12 \pm 0.02\%$. 

As a final enhancement, we increase the attack's effectiveness by accumulating evidence across multiple Gaussian sample pairs. Specifically, we use $d = 1,000$ samples $(Y_1, Y_2), \cdots, (Y_{d-1}, Y_{d})$ to make a single prediction: a candidate mean $\mu \in \{0, 1/\sqrt{d}\}$ is infeasible if
\begin{equation*}
  \bigvee_{i=1}^{d/2} \phi_\mathrm{gauss}(Y_{2i-1}, Y_{2i}; \mu, \sigma^2) = \text{\bf True}.
\end{equation*}

Note that this construction effectively describes a membership inference attack against the Gaussian mechanism $\M_\mathrm{gauss}(x) = q(x) + \mathcal{N}(0, \sigma^2\mathbb{I}_d)$, applied to vector-valued queries $q : \X \rightarrow \R^d$ with $L_2$ sensitivity of $\Delta_q = 1$.

\begin{figure}[t]
	\centering
  \includegraphics[width=\linewidth]{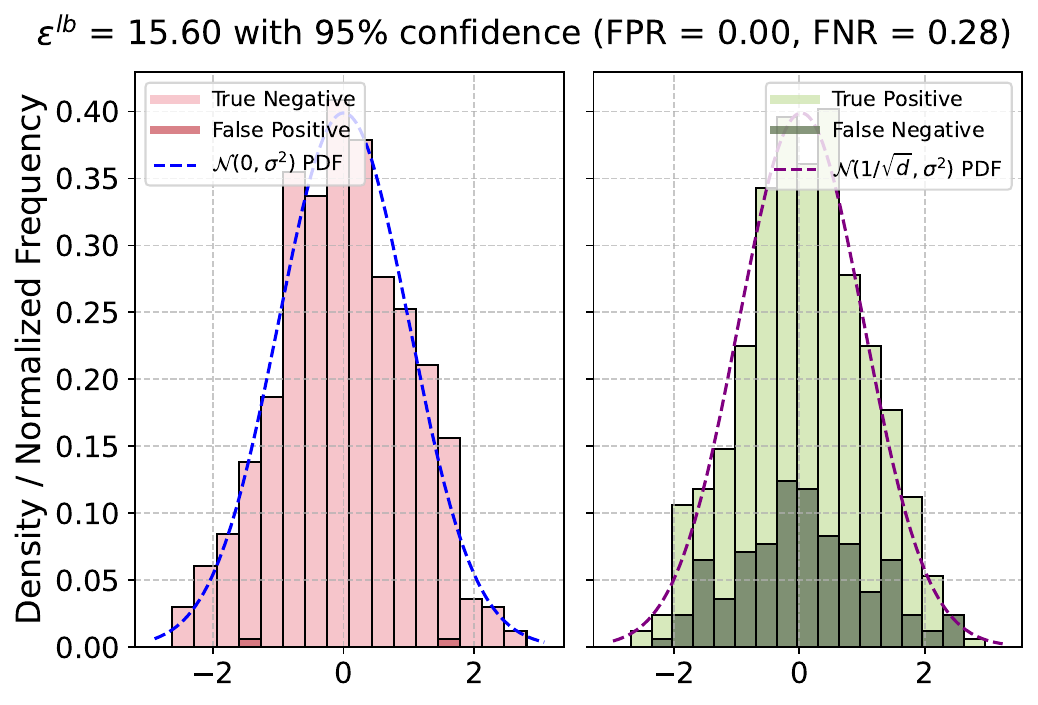}
  \caption{Result of our audit on Apple's Prio++ algorithm, which uses the $d$-dimensional Gaussian mechanism with $d=1,000$ and $\sigma^2=1$ to guarantee $(4.377,10^{-5})$-DP. The audit was conducted using the described membership test on inputs, $x_0, x_1$, such that $q(x_0)=\vec{0}$ and $q(x_1)=\vec{1}/\Vert \vec{1} \Vert$. We ran the mechanism $n=1,000$ times and the audit used the Gaussian trade-off family $\Fclass_\gaus$ and a significance level of $\sig=0.05$.}
  \label{fig:gaussian_result} 
\end{figure}

\subsubsection*{Privacy Auditing Result}
We applied this membership test to audit Apple's Prio++, which internally uses GaussianPRNG to implement the Gaussian mechanism (see~\autoref{ssec:prio_pp} for more details). As shown in~\autoref{fig:gaussian_result}, our test for distinguishing between $q(x_0) = \vec{0}$ and $q(x_1) = \vec 1 / \Vert \vec 1 \Vert$ achieves an FNR of $28.90 \pm 0.64 \%$ at $0.10 \pm 0.06\%$ FPR, resulting in an overall accuracy of $85.41 \pm 0.35 \%$. Per our audit, these results imply that the true DP parameters exceeds $\eps^\mathrm{lb} = 15.6$ for $\del=10^{-5}$. This indicates a clear DP violation since the mechanism was configured with $\sigma^2 = 1$ and $\Delta_q=1$, which should guarantee $(4.377, 10^{-5})$-DP. 


\section{Problems with Apple's Secure Aggregation}

The protection offered by DP alone does not cover many other properties that are needed to perform aggregation of statistics across billions of devices securely. For instance, DP analytics reports should be both \emph{confidential} to restrict access only to authorized servers and \emph{verifiable} to prevent malicious clients from manipulating the aggregate. In 2023, Apple announced in a research highlight~\cite{apple_differential_privacy} that it had adopted Prio~\cite{corrigan2017prio} protocol, combined with DP, to collect annotation tags and location data from users’ photos for features such as Memories and Places from iOS 17 onwards. Consistent with this, we find that Apple’s DP framework implements \hdr{_DPPrioAlgorithm.h}{Prio} and several internal variants, including \hdr{_DPPrioPlusPlusAlgorithm.h}{Prio++}, \hdr{_DPPrio3SumVectorRandomizer.h}{Prio3Sum}~\cite{ietf-ppm-dap-taskprov-03}, and \hdr{_DPPINERandomizer.h}{PINE}~\cite{rothblum2024pine}. In this section, we first outline the design and security guarantees of the original Prio~\cite{corrigan2017prio}, and then examine Apple’s implementation, highlighting key deviations that weakens their privacy.

\subsection{The Original Prio}

  Prio~\cite{corrigan2017prio} is a multiparty computation protocol for \emph{secure aggregation}, which enables a group of servers to compute an aggregate value $A = \sum_i Y_i$, without learning any individual client submission $Y_i = \M(X_i)$, where $X_i$ is the client’s private data and $\M$ privatizes it locally. Prio’s client-side protocol operates as follows.

\procb{\textbf{Client} $C_i$}{
  X_i \gets \text{Private Data} \\
  Y_i \gets \M(X_i) \pccomment{DP mechanism}\\
  ([Y_i]_1, \ldots, [Y_i]_m) \gets \secretsharesfn(Y_i;m) \\
  ([\pi_i]_1, \ldots, [\pi_i]_m) \gets \snipsfn(Y_i;m) \pccomment{ZK proof} \\
  \pcfor j = 1..m \pcdo \\
  \t\pk_j \gets \text{Server } S_j \\
  \t\msg_i^{(j)} \gets \enc(([Y_i]_j, [\pi_i]_j); \pk_j) \\
  \t\text{Send } \msg_i^{(j)} \text{ to Server } S_j
}

  Each client $C_i$ randomly splits $Y_i$ into $m$ \emph{secret shares} $[Y_i]_1, \ldots, [Y_i]_m$ over a finite field $\mathbb{F}$ such that their aggregate equals $Y_i$. Alongside the secret shares of $Y_i$, the clients also create secret shares of a proof of correctness $\pi_i$ for $Y_i$. Each pair of shares are sent to a distinct server after encryption with its public key. Prio's server-side operates as follows.
\procb{\textbf{Server} $S_j$}{
  \sk_j, \pk_j \sample \kgen \pccomment{Public key $\pk_j$ is shared with clients} \\
  \left[A\right]_j \gets \left[0\right] \pccomment{Initialized with $0$ in the field $\mathbb{F}$} \\
  \pcfor i = 1..n \pcdo \\
  \t \msg_i^{(j)}\gets C_i \\ 
  \t([Y_i]_j, [\pi_i]_j) \gets \dec(\msg_i^{(j)}; \sk_j) \\
  \t\pcif \verifyfn([Y_i]_j,[\pi_i]_j; S_1 \cdots, S_m) \pcthen \\
  \t\t \left[A\right]_j \gets [A]_j + [Y_i]_j \\
  \textbf{publish } [A]_j
}
Each server locally sums the decrypted shares it receives to obtain~$[A]_j = \sum_i [Y_i]_j$. The global aggregate is recovered by combining these partial sums as: $A = \sum_{j=1}^m [A]_j = \sum_i Y_i$.

This design ensures that as long as \emph{at least one server remains honest}, the others, even if colluding with a subset of clients, cannot reconstruct any individual~$Y_i$. In addition to being confidential, the client submissions through Prio are also \emph{verifiable}. The second set of secret-shares serve as a ZK mechanism that allows servers to collectively verify the correctness of each client’s submission, ensuring that malformed or adversarial inputs can be filtered, without even knowing the underlying values.

\subsection{Apple's implementation of Prio}
\label{ssec:apple_prio}

  The core security for MPC protocols like Prio fundamentally relies on the decentralized control of the servers performing data aggregation. When all servers are operated by the same entity in a closed-source cloud environment, the security assumption relies on operational independence between servers; from the client vantage, we cannot attest to that separation.
  In our analysis, we find that Apple's Prio implementation splits each client's data ($Y_i = \M(X_i)$) into only two secret shares~($m=2$) but transmits both to the same server endpoint, designated as the ``\emph{leader}.'' Although the shares are encrypted under distinct public keys, nominally assigning one to the ``\emph{leader}'' and other to a ``\emph{helper}'', there is no external means to verify the intended separation of decryption capabilities is upheld by Apple, effectively weakening Prio's SecAgg guarantees.

Secondly, Apple's research highlight~\cite{apple_differential_privacy} states that it employs Prio with $\eps$-DP to collect encoded photos attributes, such as approximate location and scene contents, for training models that identify iconic scenes. Consistent with this description, the \emph{keyproperties.plist} configuration file in macOS lists numerous properties, including ``\emph{InsightsIconicScenes}'', ``\emph{InsightsSafari}'', ``\emph{KeyboardTextInput}'', and ``\emph{ENDiagnosedVaccineStatus}'', as being collected via Prio. To provide $\eps$-DP, we find that Prio employs the \emph{symmetric one-hot-encoding} mechanism $\M_{\symohe}$, as described in~\autoref{alg:sym_ohe}.

\begin{algorithm}[!htb]
  \DontPrintSemicolon
  \caption{$\M_{\symohe}$~\cite[Algorithm 3]{mcmillan2022private}}\label{alg:sym_ohe}
  \KwData{Data element $X \in \{1, \cdots, d\}$, parameter $\eps$}
  \KwResult{Bit String $Y \in \{0,1\}^d$}
  Initialize $V, Y \gets \{0\}^d$\;
  $V[X] \gets 1$ \pccomment{Encode $X$ as a one-hot vector $V$}\;
  \For{$i \in \{1, \cdots, d\}$}{
    $Y[i] \gets \begin{cases}V[i] & \text{with probability } \frac{e^\eps}{e^\eps + 1} \\ 1 - V[i] & \text{otherwise} \end{cases}$\;
  }
  \textbf{Return:} $Y$\;
\end{algorithm}

\autoref{alg:sym_ohe} satisfies $\varepsilon$-DP locally, \emph{but under the deletion adjacency model}. In this model, the output distribution of $\M_{\symohe}$ on a data point~$X$ is compared against a reference distribution $\mathrm{Bernoulli}(\tfrac{e^{\eps}}{e^{\eps}+1})^d$, which corresponds to the case where nothing was sent. However, in the standard \emph{replacement adjacency model}~\cite{mcmillan2022private}, the same mechanism only achieves a much worse privacy of $2\eps$-DP.

\begin{figure}[!t]
	\centering
  \includegraphics[width=\linewidth]{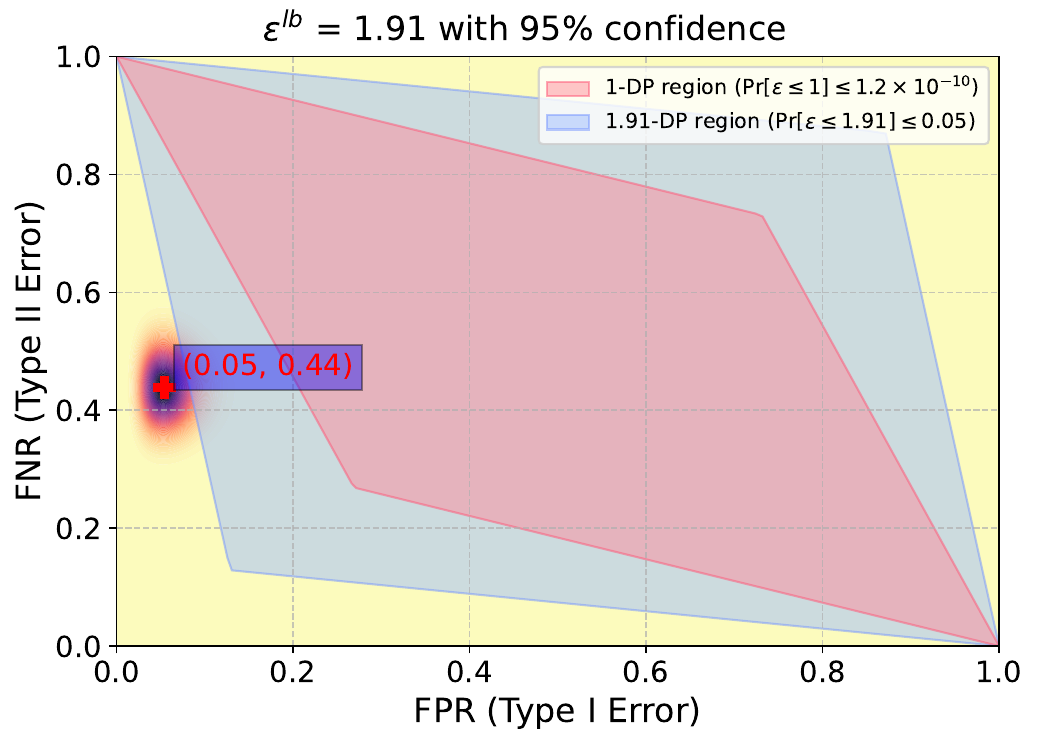}
  \caption{Illustration of our privacy audit on Apple's Prio implementation with its DP parameter set to $\eps=1$. The protocol was executed $500$ times, with the input $X$ selected randomly from $\{1, 2\}$. We then decoded the shares $([Y]_1, [Y]_2)$ to reconstruct the randomized $2$-bit vector, $Y \in \{0,1\}^2$. Our membership test predicted $X=1$ if the first bit of $Y$ was set, and as $X=2$ otherwise.}
  \label{fig:prio_audit} 
\end{figure}

A closer examination of the configuration files reveals a concerning inconsistency in how privacy budgets are defined. While most mechanisms specify $\eps$ under the replacement adjacency model, Prio appears to report its guarantees under the weaker deletion adjacency model. To verify this, we conducted a privacy audit with a membership test in which an adversary attempts to distinguish whether the input $X$ was $1$ or $2$. As shown in~\autoref{fig:prio_audit}, the observed success indicates that the actual privacy loss, i.e., the $\eps$ perceived by an adversary, is nearly twice the value configured for Prio.

Third, the configured privacy budgets for Prio are exceptionally large, ranging from $\eps = 6$ to $\eps = 8$. At such levels, a one-hot vector~$V$ with dimensionality as high as~$d = 10^4$ is almost identical to its privatized output~$Y$: they differ by at most 30 bits with $87.57\%$ probability when $\eps = 6$, and by no more than 5 bits with $87.64\%$ probability when $\eps = 8$. In other words, with such large budgets, a malicious Apple server can reconstruct sensitive data, such as domains visited on Safari or contents of pictures in the Photos app, from the collected device analytics with only a \emph{small margin of error}.

Finally, these privacy concerns extends beyond Apple servers with our discovery of \emph{unencrypted Prio client-shares publicly available on the internet}. These analytics logs 
contain raw secret-share data that can be decoded and recombined to recover the original private records. 
Such logs allow for direct reconstruction of sensitive user by an external attacker, effectively sidestepping the privacy protections that Apple's system tries to provide. ~\autoref{fig:prio_decode} shows an example of this on logs captured from our test iPhone device.

\subsection{Apple's Prio++ and PINE Variants}
\label{ssec:prio_pp}

The original Prio protocol~\cite{corrigan2017prio} was designed for verifiable and secure aggregation of \emph{discrete} data types. Its SNIP proof system for bound verification scales efficiently only for integral inputs, specifically, when the arithmetic circuit representing the proof~$\pi$ involves far fewer multiplications than the size of the underlying field~$\mathbb{F}$. Consequently, it integrates naturally with discrete DP mechanisms like~$\M_{\symohe}$. To extend its security principles to continuous domains in a computationally efficient manner, researchers at Apple have proposed protocols such as \emph{Secure Summation} (SS)~\cite{talwar2022differential} and \emph{PINE}~\cite{rothblum2024pine}, which enable verification that each client’s contribution has a bounded Euclidean norm. We find implementations of these mechanisms in the DP framework as Prio++ and PINE, respectively.

\subsubsection*{Design of Prio++}
Like Prio, Apple's Prio++ is designed to securely aggregate analytics data. However, unlike Prio, which operates on bit sequences, Prio++ handles float vectors. As a result, Prio++ employs a different mechanism for privatization (i.e., DP guarantee) and for correctness-proof generation (i.e., ZK guarantee). Specifically, Prio++ takes a float vector $X \in \R^d$ as input and applies the Gaussian mechanism to privatize it, generating
$Y = \M_\gaus(x) = X / \max(1, \Vert X \Vert) + \mathcal{N}(\vec{0}, \sigma^2\mathbb{I}_d)$.
The input $Y$ is then split into two secret shares by adding another layer of Gaussian noise with a distinct variance parameter $\sigma_\mathrm{SS}^2$ as described in~\autoref{alg:ss}.

\begin{algorithm}[!htb]
  \DontPrintSemicolon
  \caption{SecretShares$(Y)$ used in Prio++}
  \label{alg:ss}
  \KwData{Data $Y \in \R^d$, secret sharing variance $\sigma^2_\mathrm{SS}$}
  \KwResult{Leader and helper shares $[Y]_1, [Y]_2$}
  $\seedvar \sample \seedgenfn$ \pccomment{Generate a random seed}\;
  $V \sim \mathcal{N}(\vec{0},\sigma^2_\mathrm{SS}\mathbb{I}_d)$ \pccomment{By calling \hdr{_DPGaussianPRNG.h\#L16}{GaussianPRNG}(\seedvar; d)} \;
  $([Y]_1, [Y]_2) \gets (Y - V, \seedvar)$\;
  \textbf{Return:} $[Y]_1, [Y]_2$\;
\end{algorithm}

This additional noise layer serves two purposes. First, the noise layer generates two secret shares, $[Y_i]_1$ and $[Y_i]_2$, for distribution to the leader and helper server respectively. After securely aggregating their shares~$[A]_j = \sum_{i} [Y_i]_j$ from all clients, the servers combine their shares to recover the aggregate~$A = [A]_1 + [A]_2 = \sum_i Y_i$, effectively ``peeling off'' the second layer of added noise.

Second, it enables integrity verification using \emph{differentially zero-knowledge} (DZK) proofs, a relaxation of standard ZK proofs introduced in~\cite{talwar2022differential}. With the two shares, $[Y_i]_1$ and $[Y_i]_2$, servers can probabilistically verify that $\Vert Y_i \Vert \leq \uprho$ for some~$\uprho > 1$ before accepting a client's submission, without revealing excessive information to each other (see~\cite[Algorithm 3]{talwar2022differential} for verification details).

\subsubsection*{Violation of DZK}
The floating-point vulnerabilities found in GaussianPRNG---which we showed in \autoref{sssec:prio} cause the Gaussian mechanism used in Prio++ to violate its DP guarantees---also affect the SecretShares$(Y)$ function (in \autoref{alg:ss}). Because it uses the same flawed PRNG to generate shares, the resulting leader's share, $[Y]_1 = Y - \mathcal{N}(\vec{0}, \sigma_\mathrm{SS}^2\mathbb{I}_d)$, is not nearly as indistinguishable to $\mathcal{N}(\vec{0}, \sigma_\mathrm{SS}^2\mathbb{I}_d)$ as necessary for the DZK property to hold. This DZK failure is a direct consequence of the shared vulnerability, and we demonstrate it using a membership-inference attack like the one in \autoref{sssec:prio}. This vulnerability also affects the DZK property of Prio++Metrics.

\subsection{Apple Disables DP in SecAgg Protocols}

By design, secure aggregation is intended to serve as an \emph{additional layer of security} on top of the local DP applied to the device analytics---for sufficiently small~$\eps$, this local $\eps$-DP guarantee is amplified to $(O(\eps\log(1/\del) / \sqrt{n}), \del)$-DP centrally when aggregating submissions from $n$ users~\cite{mcmillan2022private}. An intention to use this synergy is evident in Apple's implementations as all three protocols---Prio, Prio++, and PINE---include options to enable local DP, either through a \hdr{_DPPrioAlgorithm.h\#L27}{\emph{dp}} flag or directly through a configuration \hdr{_DPPINERandomizer.h\#L32}{\emph{metadata}}.

However, we discovered that these protocols are \emph{configured to disable the DP layer}
for the vast majority of collected analytics data. Furthermore, we discover that sub-variants such as Prio++Metadata and Prio++Metrics are essentially wrappers around Prio and Prio++ that \emph{hard-code DP to be disabled}. This means that data collected under these variants \emph{lacks local differential privacy guarantees}.
Our DP audit confirmed this alarming operation as our membership inference attacks achieved a $100\%$ accuracy rate. Furthermore, Apple’s response to our report acknowledges that these protocols \emph{intentionally lack} DP guarantees  (cf. {\def\subsectionautorefname{Appendix}\autoref{ssec:apple_confirms_no_dp}}). In total, we find that at least $40\%$ of the collected data in Sonoma 14.2 and $37\%$ in Sequoia 15.6, including sensitive properties like ``\emph{QuickTypeDESPlugin.LSTM}'', ``\emph{NightangleTraining}'', ``\emph{Siri.GlobalSpeechNNLM}'', and ``\emph{IntelligencePlatform}'', are collected with DP disabled.

\begin{table}[!tb]
\centering
\caption{Summary of discovered issues in SecAgg protocols}
\label{tab:prios}
\begin{tabular}{cccc}
\toprule
\textbf{Protocol} & \textbf{DP Mech.} & \textbf{PRNG for SecretShares} & \textbf{ZK} (\cmark/\xmark) \\
\midrule
\multicolumn{4}{c}{\textit{Violates DP under the replacement adjacency model}} \\
Prio & $\M_{\symohe}$ & PrioSeedablePRNG & ZK \cmark \\
\midrule
\multicolumn{4}{c}{\textit{Violates DP due to floating-point vulnerabilities}} \\
Prio++ & $\M_\gaus$ & GaussianPRNG & DZK \xmark \\
\midrule
\multicolumn{4}{c}{\textit{Does not implement DP}} \\
Prio++Metadata & \textbf{None} & PrioSeedablePRNG & ZK \cmark \\
Prio++Metrics & \textbf{None} & GaussianPRNG & DZK \xmark \\
\bottomrule
\end{tabular}
\end{table}

\subsection{Decoding Prio and Prio++ Logs in iPhones}
\label{sssec:decodeprio}

\begin{figure}[!t]
	\centering
  \includegraphics[width=\linewidth]{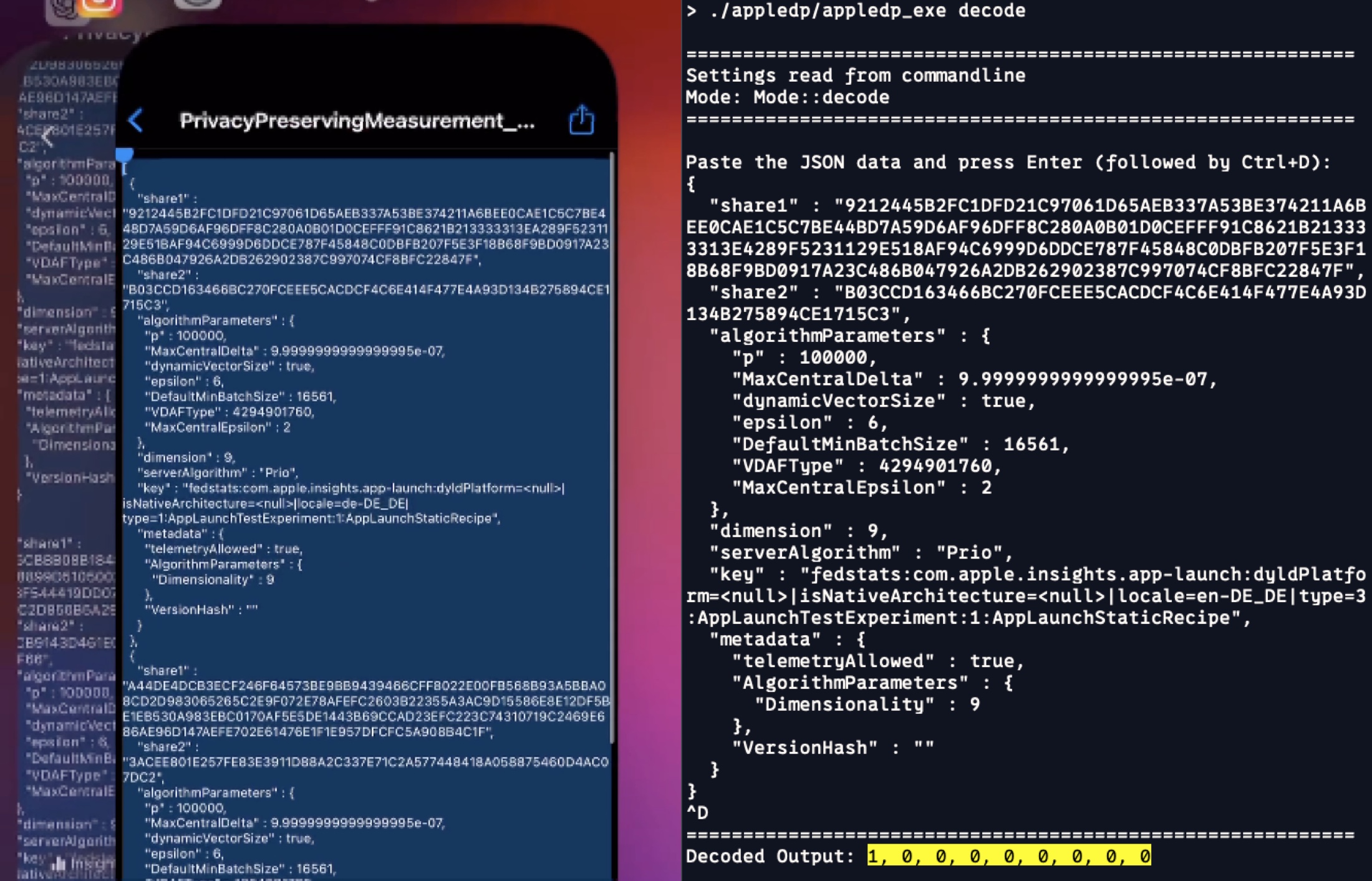}
  \caption{Decoding analytics data from our test iPhone device. The logs reveal unencrypted `\emph{share1}' and `\emph{share2}' fields (representing the secret shares $[Y]_1$ and $[Y]_2$) of the differentially private measurement $Y$ uploaded to Apple's servers. 
  These plaintext logs allow direct decoding of $Y$, as shown on the right. Furthermore, the logs indicate the original OHE of $X$ had a dimensionality of $9$, privatized with $\eps\!=\!6$, leading to a $97.80\%$ chance that the decoded $Y$ equals the original $X$.}
  \label{fig:prio_decode} 
\end{figure}

Apple's 2017 white paper introducing its DP framework~\cite{apple2017learning} stated that analytics logs generated by the system would be visible on iOS under Settings $\rangle$ Privacy \& Security $\rangle$ Analytics \& Improvements $\rangle$ Analytics Data. Our investigation finds this to be \emph{only partially true} in 2025; over $68\%$ of the collected properties in macOS Sonoma 14.2 are configured with `\emph{DirectUpload~$=$~true}', meaning they are transmitted directly to Apple's servers without generating any local logs. Nevertheless, we confirmed that a subset of analytics data continues to be logged locally; over three months of observation on an iPhone~12 running iOS~18.1.1, we repeatedly found analytics entries prefixed with `\emph{PrivacyPreservingMeasurements}', as shown in \autoref{fig:prio_decode}.

Examining these logs, `\emph{share1}' and `\emph{share2}' attributes stand out.
Our analysis of the SecretShares$(Y)$ implementation in both protocols reveals that these fields directly contain the \emph{leader} and \emph{helper} secret shares in \emph{plaintext}---without any encryption applied. Combined with the `\emph{algorithmParameters}' attribute present in the same logs, these three fields are sufficient to fully decode and reconstruct the original private data. 

Specifically, Prio employs a SecretShares$(Y)$ function similar to \autoref{alg:ss}, but replaces GaussianPRNG with PrioSeedablePRNG and appends a ZK proof, as described in~\autoref{alg:ss_p}.
\begin{algorithm}[!htb]
  \DontPrintSemicolon
  \caption{SecretShares$(Y)$ used in Prio}
  \label{alg:ss_p}
  \KwData{Data $Y \in \{0,1\}^d$}
  \KwResult{Leader and Helper shares $[Y]_1, [Y]_2$}
  $\seedvar \sample \seedgenfn$ \pccomment{Generate a random seed}\;
  $V \sim [0, 2^{32} - 1]$ \pccomment{By calling \hdr{_DPPrioSeedablePRNG.h}{PrioSeedablePRNG}(\seedvar; d)}\;
  $[Y]_1 \gets (Y - V,  \snipfn(Y, \seedvar)) $ \pccomment{\shareonefield}\;
  $[Y]_2 \gets \seedvar$ \pccomment{\sharetwofield} \;
  \textbf{Return:} $[Y]_1, [Y]_2$\;
\end{algorithm}
To reconstruct the original data, we invert \autoref{alg:ss} and \autoref{alg:ss_p} by computing 
$Y \gets [Y]_1+\prngfn([Y]_2)$
where $\prngfn$ is PrioSeedablePRNG for Prio (and its child Prio++Metadata), and GaussianPRNG for Prio++ (and Prio++Metrics). For Prio, the addition is performed in the finite field~$\mathbb{F}$ over `\emph{uint32}', while for Prio++, it is computed in `\emph{float32}'.
We verified that this decoding process works across all four Prio variants, enabling nearly full recovery of the underlying private data directly from the analytics logs generated in our test iPhone device, as shown in \autoref{fig:prio_decode}.

\section{Use of Large DP Parameters}

We also audit several other DP mechanisms within Apple’s framework that sample from discrete distributions. Such mechanisms are simpler to implement with PRNGs and less likely to violate differential privacy in practice. While our audit did not find any DP violations, we highlight that their configured privacy budgets are large relative to common practice and can weaken privacy.

\subsection{Count Sketch Mechanisms}
\label{ssec:cms}

Apple's 2017 white paper proposed the \hdr{_DPCMSRandomizer.h}{Count Median Sketch (CMS)} and \hdr{_DPHCMSRandomizer.h}{Hadamard Count Median Sketch (HCMS)} algorithms~\cite{apple2017learning}, which we find directly implemented within the DP framework. These algorithms operate by collecting randomized encodings of count data from the users to estimate the frequencies of a set of known elements, such as emojis used in QuickType and web domains visited on Safari. At the server, these encodings are then aggregated into a \emph{count-sketch matrix}, which can then be queried to extract a privacy-preserving approximate count of all the elements. \autoref{alg:cms} and \autoref{alg:hcms} detail the client-side part of the CMS and HCMS algorithms respectively.
\begin{algorithm}[!htb]
  \DontPrintSemicolon
  \caption{$\M_{\clientcms}$~\cite[Algorithm 2]{apple2017learning}}\label{alg:cms}
  \KwData{Data element $X$, parameter $\eps$, bit count $d$, hash count $k$}
  \KwResult{Bits $Y \in \{0,1\}^d$, hash index $j$}
  Uniformly sample $j \sim [k]$\;
  Initialize vector $V \gets \{0\}^d$\;
  Set $V[\hash_j(x) \mod d] \gets 1$\;
  \For{$i \in \{1, \cdots, d\}$}{
    $Y[i] \gets \begin{cases}V[i] & \text{with probability } \frac{e^{\eps/2}}{e^{\eps/2} + 1} \\ 1 - V[i] & \text{otherwise} \end{cases}$\;
  }
  \textbf{Return:} $Y$, index $j$\;
\end{algorithm}

The mechanism $\M_{\clientcms}$ works by hashing the input $X$ into one of $d$ possible buckets using a random hash function $\hash_j$, implemented as $\shafn(j \Vert x)$ in the DP framework, to create a one-hot vector $V$. This vector is then privatized with $\eps$-DP by randomly flipping its bits and then sent to the server along with the hash function index $j$. 
\begin{algorithm}[!htb]
  \DontPrintSemicolon
  \caption{$\M_{\clienthcms}$~\cite[Algorithm 6]{apple2017learning}}\label{alg:hcms}
  \KwData{Data element $X$, parameter $\eps$, bit count $d$, hash count $k$}
  \KwResult{Bit $Y \in \{-1,1\}$, hash index $j$, bit index $l$}
  Uniformly sample $j \sim [k]$\;
  Initialize vector $V \gets \{0\}^d$\;
  Set $V[\hash(j,x) \mod d] \gets 1$\;
  Transform $U \gets H_d \times V$ \pccomment{$H_d$ is the Hadamard matrix} \;
  Uniformly sample $l \sim [d]$\;
  Set $Y \gets \begin{cases}U_l & \text{with probability } \frac{e^{\eps}}{e^{\eps} + 1} \\ -U_l & \text{otherwise} \end{cases}$\;
  \textbf{Return:} $Y$, index $j$, index $l$\;
\end{algorithm}

The $\M_{\clienthcms}$ is similar to $\M_{\clientcms}$, except it compresses the one-hot vector $V$ by multiplying it with the Hadamard matrix $H_d$ and then randomly selecting an index $l \in [d]$ of the transformed vector $U$ to privatize with $\eps$-DP and send to the server, along with the indices $l$ and $j$.

\begin{figure}[!t]
	\centering
  \includegraphics[width=\linewidth]{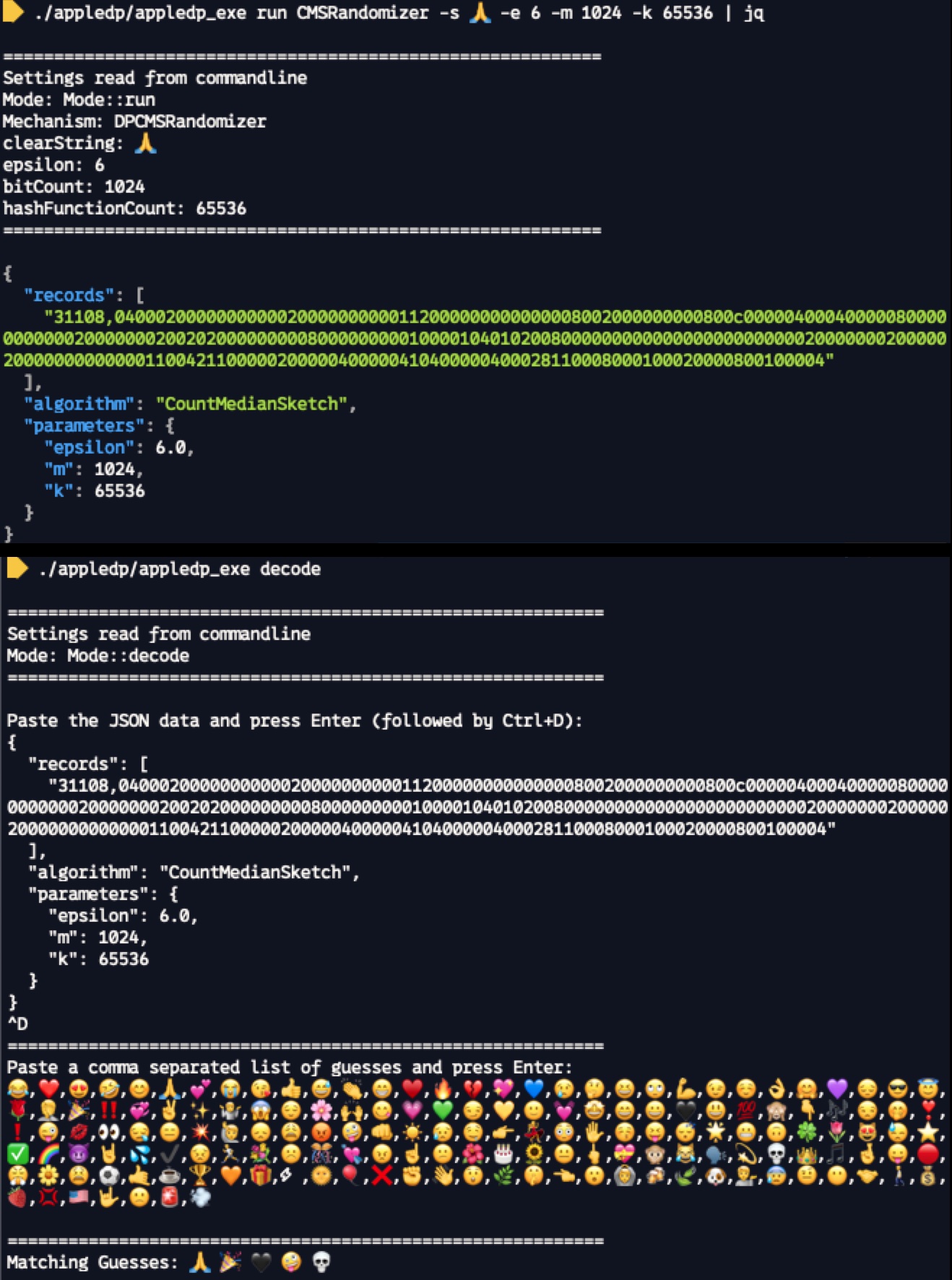}
  \caption{Illustration of our decoder for Apple's Count Median Sketch ($\M_{\clientcms}$). Our decoder's functionality is demonstrated by a test case: the emoji ``\prayer'' is privatized using $\M_{\clientcms}$ with a privacy budget of $\eps=6$. We then attempt to identify the original emoji from a predefined set of 152 most popular emojis. The decoder successfully narrows down this set to 5 potential emojis, accurately including the true input.}\label{fig:our_cms} 
\end{figure}

\subsubsection*{Decoder for $\M_{\clientcms}$ and $\M_{\clienthcms}$}
We found several leaked logs from Apple's CMS and HCMS mechanisms, containing keynames like ``\emph{keyboard.Emoji.en\_US}'' or ``\emph{PodcastTopic.en\_US}'' that describe the type of recorded data. To avoid any risk of exposing real user data, we demonstrate our decoders on outputs that we generated ourselves using custom executables that call $\M_{\clientcms}$ and $\M_{\clienthcms}$ mechanisms within Apple's DP framework (see~\autoref{fig:our_cms}).

Our decoding approach requires a candidate set $G$ of plausible values for the private data $X$ (a set that, in principle, could be formulated for any given log based on its keyname). For a given hash index $j$ and corrresponding output vector $Y$, we consider a guess $g \in G$ plausible if hasing it yields an index whose bit in $Y$ is set, i.e., $Y[\hash_j(g) \mod d] = 1$. The rationale behind this is that when $\eps \geq 4$, there is at least an $88.08\%$ probability that the true bit in the one-hot vector $V$ remains unflipped in the privatized output $Y$. Hence, any $g \in G$ that maps to a set bit in $Y$ is a plausible candidate for the original value of $X$. 

For $\M_{\clienthcms}$, we apply a similar decoding process that accounts for the Hadamard transformation. Specifically, for an observation $(Y, j, l)$ from $\M_{\clienthcms}$, we mark $g \in G$ as plausible if hasing it with $\hash_j$ yields a transformed vector whose bit at index $l$ matches the observed output $Y$, i.e., $\ohefn(l)^\top \times H_d \times \ohefn(\hash_j(g) \mod d) = Y$.

\subsubsection*{Reconstruction Attack}
\begin{figure}[!t]
	\centering
  \includegraphics[width=\linewidth]{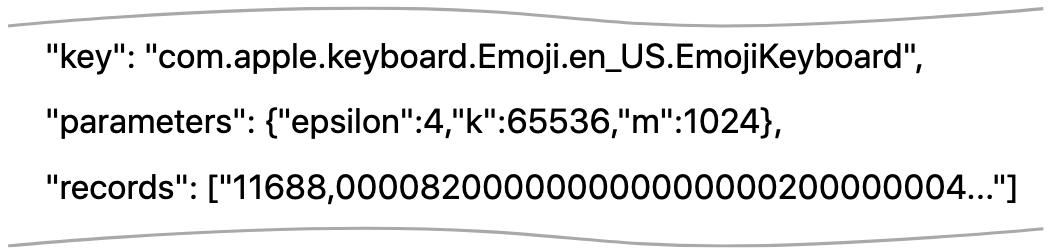}
  \caption{Sample record generated by Count Median Sketch that appears in~\cite[Figure 3]{apple2017learning}. Running our decoder on it after padding ``...'' with $0$s on the same list of 152 emojis as in \autoref{fig:our_cms} results in a unique match: ``\laugh''.}\label{fig:apple_paper_cms} 
\end{figure}

We find that our strategy is quite effective in decoding the output of $\M_{\clientcms}$, especially when provided with a descent set of guesses $G$ for the target $X$, as illustrated in \autoref{fig:our_cms}. 
When we tested our decoder on outputs generated from our executable---configured to mimic the leaked logs with the keyname ``\emph{keyboard.Emoji.en\_US}'' and with $\eps=6$ budget---it frequently returned just a single matching emoji. Moreover, when we ran the decoder on the partial CMS sample from Apple’s white paper (cf. \autoref{fig:apple_paper_cms}), it returned a unique match: ``\laugh''. Given that the paper itself reports ``\laugh'' as the overwhelmingly most popular emoji in the English locale (see~\cite[Figure~5]{apple2017learning}), this provides strong evidence that our reconstruction was successful, illustrating the privacy risks.

\subsubsection*{Privacy Auditing Result}
We conducted a DP audit of $\M_{\clientcms}$ and $\M_{\clienthcms}$ using a membership test grounded in our decoding strategy. In this setup, the adversary predicts the input as $x_1 =$ ``\greentick'' if the alternative guess $x_0 =$ ``\redcross'' fails to decode to the mechanism’s output. Each mechanism was executed $n = 1000$ times with $\eps = 4$, randomly selecting between $x_0$ and $x_1$ in each run. From these experiments, we obtained $95\%$-confidence lower bounds of $\eps^\mathrm{lb} = 2.86$ for $\M_{\clientcms}$ and $\eps^\mathrm{lb} = 3.66$ for $\M_{\clienthcms}$ (see~\autoref{alg:zb_auditor}), confirming that both adhere to their stated privacy parameters.
However, the membership inference attacks achieved high success rates---83.80\% accuracy (FPR $2.65\%$, TPR $70.73\%$) for $\M_{\clientcms}$ and $71.30\%$ accuracy (FPR $0.41\%$, TPR $44.01\%$) for $\M_{\clienthcms}$. These results show that, even with a correct implementation, a privacy budget of $\eps = 4$, which Apple used from 2017 to 2022 for collecting sensitive data (see {\def\subsectionautorefname{Appendix}\autoref{ssec:historical_mechanisms}}), was far too large to provide meaningful privacy protection.

\subsection{PI-RAPPOR Mechanism} 
\label{ssec:pirappor}

Another mechanism that we audit is PI-RAPPOR. Introduced in \cite[Algorithm 3]{feldman2021lossless}, this mechanism is configured with constants $0 < \alpha_0 < \alpha_1 < 1$, number of classes $k$, and a prime $p$ such that $\alpha_0 p \in \mathbb{N}$, and satisfies $\eps$-DP in the standard replacement model for $\eps = \log \frac{\alpha_1(1-\alpha_0)}{\alpha_0(1-\alpha_1)}$. The output of this mechanism is a pair of coefficients, $(\phi_0, \phi_1)$, that is transmitted to the servers. For brevity, we omit the description of this mechanism, referring the reader to~\cite[Section 4.1]{feldman2021lossless} for the details of its design.

We observe that the implementation of PI-RAPPOR closely aligns with its specification in~\cite[Algorithm 3]{feldman2021lossless}. Its constructor takes the same four initialization parameters, which we set to $k=2$, $\alpha_0=0.000335$, $\alpha_1=0.999665$, and $p=2971$, yielding $\eps$-DP with $\eps=8$. These values correspond exactly to those used in macOS Sonoma 14.2 for collecting analytics data associated with keys like ``\emph{systemsettings.SearchTerm}'', ``\emph{networking.IPPrefix}'', ``\emph{keyboard.WordUnigram}'', and others. 

To decode an output $(\phi_0, \phi_1)$, we employ the built-in \hdr{_DPPiRapporAlgorithm.h/\#L28}{\emph{decode()}} method provided in the DP framework. Using this decoder as the membership test for our privacy auditing (via~\autoref{alg:zb_auditor}), we evaluated PI-RAPPOR on inputs $x_0 = 0$ and $x_1 = 1$ across $n = 1,000$ runs, obtaining a $95\%$-confidence lower bound of $\eps^\mathrm{lb} = 5.51$. Although our audit does not indicate any DP violation, we note that the configured privacy budget of $\eps = 8$ is nonetheless far too large to provide sufficient protection, as our membership inference attack still achieved a striking $98.43\%$ accuracy.

\subsection{OneBitHistogram Mechanism} 
\label{ssec:obh}
The final mechanism we audit in Apple's DP framework is OneBitHistogram, which also seems to be designed for the frequency estimation task. Although we could not find any public information from Apple on its design, OneBitHistogram seems to be a variant of the Randomized Response mechanism that resembles~\cite[Algorithm 1]{bassily2015local}. Based on our inspection of the binary, we describe it in~\autoref{alg:obh}.

\begin{algorithm}[!h]
  \DontPrintSemicolon
  \caption{OneBitHistogram}\label{alg:obh}
  \KwData{Data element $X$, parameter $\eps$, bit count $d$}
  \KwResult{Bit $Y \in \{0,1\}$, bit index $l$}
  Uniformly sample index $l \sim [d]$\;
  Create key $\key \gets \shafn(X)$\;
  Encrypt $V \gets \aesfn(l;\key)$\;
  $Y \gets \begin{cases} V[l] & \text{with probability} \frac{e^\eps}{e^\eps + 1}\\ 1 - V[l] &\text{otherwise} \end{cases}$\;
  \textbf{Return:} Bit $Y$, bit index $l$\;
\end{algorithm}

We do not know why \autoref{alg:obh} was designed in its presented form, nor how Apple’s servers aggregate the data it produces. Nevertheless, given its output $(Y, l)$, we can test whether a guess $g \in G$ for the true data element $X$ is plausible by checking whether $\aesfn(l;\shafn(g))[l] = Y$. Using this membership test to distinguish between the guesses $G = \{\text{``\redcross"}, \text{``\greentick"}\}$, we audit the DP of the OneBitHistogram mechanism (via \autoref{alg:zb_auditor} in {\def\sectionautorefname{Appendix}\autoref{sec:app_audit}}). With $n = 1{,}000$ runs, we obtain a 95\%-confidence lower bound of $\eps^\mathrm{lb} = 0.86$, when the mechanism was configured to satisfy $1$-DP. While we did not detect any violations of DP, our membership attack achieves an accuracy of $72.10\%$ (FPR $28.92\%$, TPR $73.08\%$), which is strikingly close to the theoretical bound $e^\eps/(e^\eps+1) \approx 73.11\%$ for $\eps = 1$.


\section{Countermeasures and Discussion}
\label{sec:counter}

Our audit of Apple's DifferentialPrivacy.framework reveals a significant gap between the company's public commitment to privacy and the technical reality of its implementation. The vulnerabilities we uncovered---ranging from flawed noising mechanisms to insecure multiparty computation protocols and excessively weak privacy parameters---undermine the very guarantees that differential privacy is meant to provide. These findings not only impact user privacy on potentially billions of devices but also highlight systemic issues in deploying privacy-enhancing technologies at scale without sufficient transparency. Here, we propose concrete countermeasures to address the identified flaws and discuss the broader implications for Apple and the industry.

\subsection{Mitigating Implementation Vulnerabilities}

The vulnerabilities detailed in this paper are not theoretical; they are practical flaws with tangible impacts on user privacy. Fortunately, they are also rectifiable. We propose the following technical countermeasures.

\subsubsection*{Secure Noise Generation}
The use of floating-point arithmetic for generating noise from continuous distributions like Laplace and Gaussian is a well-documented anti-pattern in DP implementation~\cite{mironov2012significance,jin2022we,lokna2023group}. We recommend that Apple deprecates its current noise generation mechanisms based on inverse transform and the Marsaglia polar methods. Instead, it should adopt SOTA, secure techniques designed to be resilient against floating-point attacks. Options include switching to discrete versions of these distributions (e.g., the Discrete Laplace or Discrete Gaussian distributions~\cite{canonne2020discrete}) or employing a secure noise library or other provably safe methods~\cite{google_secure_noise,holohan2024securing,keller2024secure}. This would directly remediate the flaws found in NumberRandomizer and Prio++.

\subsubsection*{Re-Architecting Secure Aggregation Protocols}
Our analysis shows that Apple’s implementation of the Prio~\cite{corrigan2017prio}-based secure aggregation protocol has several issues that weaken the privacy it provides to its users. We provide the following recommendations to address them:

\begin{enumerate}[label=\textbf{\arabic*})]

    \item \textbf{A Public Server.} A SecAgg protocol's security guarantee is predicated on the operational independence of the servers aggregating the secret shares. Sending both shares to a single endpoint falls outside standard threat‑model assumptions about independently operated servers. To rectify this, the system should be re-designed to send shares to \emph{two or more logically and physically separate servers}. Critically, to ensure public trust, one of these aggregation servers should be a \emph{publicly auditable entity}, allowing external parties to verify that the protocol is being followed correctly.

    \item \textbf{Always use DP.} All data transmitted, including metadata and performance metrics, should be considered potentially sensitive. Protocols like Prio++Metadata and Prio++Metrics currently transmit information without any DP noise, relying only on SecAgg for protection, which is problematic. All components of an analytics data should be protected with a formal differential privacy guarantee, ensuring that no part of the collected data can be used to compromise user privacy.

    \item \textbf{Cryptographically secure PRNGs.} The random numbers used to generate the shares must come from a cryptographically-secure PRNGs. Re-using the vulnerable floating-point-based GaussianPRNG for secret sharing, as is done in Prio++, Prio++Metrics and PINE, should be avoided.

    \item \textbf{Encrypted Logs.} \chgrep{The current practice of logging private records' in plaintext is a critical vulnerability, especially given that these logs often leak online. Each privatized record should be encrypted immediately upon generation, and they should appear only in their encrypted form within any analytics logs visible on the user's device.}{The current practice of exposing private record payloads in user-exportable diagnostics is a critical vulnerability, especially once such logs leave the device. The surrounding metadata can remain available for debugging, but sensitive payload fields such as secret shares and encoded measurements should be encrypted whenever reports are staged for export or user sharing.}

    \item \textbf{Preventing Side-channel Attacks.} The very act of generating a report or uploading analytics data can leak information, such as how often a user engages with a particular app. To curb this, the frequency and timing of report generation must be randomized with a DP mechanism to obscure the underlying activity patterns.

\end{enumerate}

\subsubsection*{Tighter Privacy Budgeting}
Even a perfectly implemented DP algorithm can offer negligible privacy if configured with a large privacy budget $\eps$. Apple should conduct a thorough review of the privacy budgets assigned to all the collected properties. Any property with an $\eps$ value high enough to permit trivial membership inference (e.g., $\eps > 4$) should be considered effectively non-private. These budgets must be drastically lowered or their use explicitly justified with clear, user-facing consent. Additionally, the ambiguity we observed in the privacy budget for Prio---where the configured $\eps$ appears to correspond to a $2\eps$ loss when doing membership inference---highlights the need for clear and consistent definitions of neighbouring relationship in privacy guarantees across all mechanisms.

\subsection{Transparency and Verifiability}

While technical fixes are essential, our findings point to a deeper, more structural problem: the opacity of Apple's privacy engineering. Relying on ``\emph{security through obscurity}'' is a fragile strategy, especially for primitives like DP where subtle implementation errors can silently break privacy guarantees. Our DP audit validates long-held concerns from the privacy community regarding Apple's closed-source approach~\cite{Cyphers2017,HackerNews2016}. 
Slogans like ``\emph{what happens on your iPhone, stays on your iPhone}'' ring hollow if the mechanisms ensuring this are difficult to verify and, as we have shown with our privacy audits, riddled with flaws. 

The most effective measure to restore trust on its privacy features is \emph{transparency}. We recommend that Apple \emph{open‑source, or publish a verifiable specification of, the DifferentialPrivacy.framework library}.
Doing so would allow for public verification from independent researchers and security experts, fostering a collaborative environment to identify and fix vulnerabilities before they can be exploited.


\section{LLM Usage Considerations}

We leveraged LLMs, specifically ChatGPT 4o and Gemini 2.5 Pro models, in understanding complex sections of the decompiled binary. These LLMs were also used for editorial purposes in this manuscript, and all outputs were inspected by the authors to ensure accuracy and originality.

\section{Ethics Considerations}
\label{sec:ethics}

All experiments illustrated in this work used only the author‑controlled test devices and synthetic inputs. Throughout the duration of this study, no third‑party user data was collected, retained, or analyzed. 

The objective of our ethical reverse-engineering efforts was to responsibly disclose any vulnerabilities we identify to Apple directly. We did so on August 13, 2025, submitting an early draft of this work, accompanying proof-of-concept exploits, and references to publicly exposed logs that appeared susceptible to data extraction attacks. Apple responded promptly the following day, requesting clarification on our publication timeline. By August 29, 2025, the company confirmed \chgadd{in its disclosure correspondence with us} that the reported vulnerabilities were reproducible and stated that fixes were planned for release in fall 2025\chgdel{, as shown in the disclosure screenshot below}.
\chgdeletefigure{fig:apple_initial_response}{screenshot of Apple’s initial responsible-disclosure acknowledgment}{Apple’s initial acknowledgment and commitment to address the reported issues.}

However, two months later, Apple reclassified the vulnerabilities as ``\emph{expected behaviour}'' and closed the report on October~7, 2025. {\def\sectionautorefname{Appendix}\autoref{sec:apple_response}} summarizes Apple's official response that we received at the time of closure of our vulnerability report, and explains its relation to our results.

\section*{Acknowledgments}

This research is supported by the Ministry of Education, Singapore, under an AcRF Tier 2 Grant (MOE-000761-01), and by a gift grant from Google.


\bibliographystyle{IEEEtran}
\bibliography{references.bib}


\appendices

\section{Apple’s Response and Relation to Our Results} \label{sec:apple_response}

This appendix summarizes Apple’s response to our disclosure and explains how they relate to the results that we report. We note that all results in the paper concern the \emph{client-side, pre-aggregation} behavior of the mechanisms as shipped in macOS Sonoma 14.2 and Sequoia 15.6. We did not instrument or evaluate Apple’s server-side systems. 

\subsection{Deployment Status and Versions}
\label{ssec:apple_deployment_status}

Apple states that several mechanisms that we examined (e.g., the Laplace sampler in NumberRandomizer, GaussianPRNG-based Prio++ paths, and CMS/HCMS variants) have been \emph{deprecated} and that newer deployments use PINE that implements open protocol designs like \emph{Verifiable Distributed Aggregation Functions} (VDAF)~\cite{irtf-cfrg-vdaf-17} and \emph{Distribution Aggregation Protocol} (DAP)~\cite{ietf-ppm-dap-16}. Our results therefore pertain to the client code and configurations as shipped in macOS Sonoma 14.2 and Sequoia 15.6, in which many of the now-deprecated mechanisms were still in active use. 

\subsection{Floating-Point Sampling and PINE}

Apple states that PINE converts outputs to integers before secret sharing. Our audits and reconstruction tests target the Laplace and Gaussian pathways that we executed via dynamic loading; we report violations for NumberRandomizer and for the Gaussian pathway used by Prio++. Although we found that \hdr{_DPPINERandomizer.h\#L29}{PINE's addNoiseToData()} method also uses the vulnerable GaussianPRNG, we did not evaluate its \hdr{_DPPINERandomizer.h\#L34}{randomizeFloatVector()} method and therefore do not make claims about its privacy or confidentiality guarantees.

\subsection{Leader/Helper Separation}

Apple states that the leader forwards one share to a distinct helper holding a different key. From the client side, we observe uploads of two ciphertexts to a leader endpoint; we cannot externally attest to operational non‑collusion or infrastructure separation -- this is a trust assumption not auditable from the client vantage. Our findings about \emph{pre‑aggregation} risk are unaffected.

\subsection{Configurations Collected Without Local DP}
\label{ssec:apple_confirms_no_dp}

Apple confirms that certain variants (e.g., PINEMetadata and PINEMetrics) are collected \emph{without} local DP and rely on secure aggregation. Our results likewise found local‑DP‑off configurations (previously Prio++Metadata and Prio++Metrics) and showed perfect membership inference pre‑aggregation under those settings. We describe these precisely as ``configured with local DP disabled'' and retain the analysis of risk to any party that obtains client‑side logs.

\subsection{Plaintext Logs and Reconstruction Risks}

Apple notes that on‑device analytics logs are plaintext and argues access requires an unlocked device; logs in transit are encrypted. Our result targets a \chgrep{different}{different, post-export} threat model: \chgadd{we do not claim that arbitrary third-party apps can read these files directly on-device under Apple's sandbox. Rather, }any party (not necessarily Apple) that obtains pre‑aggregation logs (e.g., via user export, diagnostics sharing, or online leaks) can reconstruct measurements from `\emph{share1}', `\emph{share2}', and recorded parameters. This does \emph{not} imply that Apple’s servers see decrypted per‑user data; it shows pre‑aggregation exposure to parties with log access.

\subsection{Count–Sketch Mechanisms}
\label{ssec:historical_mechanisms}

Apple reports that Count-sketching mechanisms have been deprecated and have not used for data collections since 2022. Matching this, we did find CMS/HCMS mechanisms missing in the DP framework shipped on macOS Sequoia 15.6 (but were present in Sonoma 14.2). We keep our results for these mechanisms as \emph{historical} to illustrate how large $\varepsilon$ materially weakens privacy, because other deployers may still use these designs. Our decoders and audits demonstrate the methodology without relying on third‑party logs.

\subsection{Adjacency Models and Shuffling}

Apple emphasizes privacy amplification by shuffling, justifying the use of large Local DP budgets. Our audits estimate \emph{local, pre‑shuffle} privacy (the view observable at the client and in the logs). Shuffling can strengthen central DP at the aggregate, but it does not mitigate \emph{log‑leak} risk because an attacker holding pre‑aggregation logs sees the local outputs before shuffling. 

Apple also emphasizes deletion adjacency model as being widely trusted and used in the industry. Our analysis highlights the inconsistency in how adjacency is applied: whereas the other mechanisms we audited interpret their `\emph{PrivacyParameter}' in the configuration under replacement adjacency, Prio interprets it under deletion adjacency.

\subsection{Zero-Knowledge Proofs}

Apple states that PINE’s zero-knowledge proofs are unaffected. We make no claim about PINE’s ZK proofs. Our DZK‑related finding pertains to Prio++, where the same floating‑point issue that affects the Gaussian mechanism also impacts the SecretShares generator and thereby the indistinguishability requirement used in its DZK verification.

\section{Our DP Auditing Framework}
\label{sec:app_audit}

We consider a mechanism $\M: \X \rightarrow \mathrm{Prob}(\Y)$ that takes as input a record from the data universe $\X$ and produces a random outcome in the output space $\Y$.
\begin{definition}[Differential Privacy~\cite{dwork2019differential}]
  An mechanism $\M$ is $(\eps, \del)$-\emph{differentially private} ($(\eps, \del)$-DP) if for all $\x, \x' \in \X$, with the output distributions $\M(\x), \M(\x')$ denoted as $P, Q$ respectively, the following condition holds:
  \begin{equation}
    \forall S \subset \Y \ : \ P(S) \leq e^\eps \cdot Q(S) + \del.
  \end{equation}
\end{definition}
A practical interpretation of DP guarantees is to see them as constraints on a \emph{hypothesis test}---let $y \in \Y$ be an output obtained by running the mechanism $\M$ on either $\x$ or its neighbour $\x'$, then any membership test $\phi: \Y \rightarrow \{0, 1\}$ to distinguish between the \emph{null} and \emph{alternate} hypothesis:
\begin{align*}
  H_0 : y\ \text{came from}\ \M(\x), \quad
  H_1 : y\ \text{came from}\ \M(\x'), 
\end{align*}
cannot simultaneously have a low \emph{false alarm rate} $\fpr_\phi$ (i.e., Type-I error or False Positive Rate or FPR) and a low \emph{missed detection rate} $\fnr_\phi$ (i.e. Type-II error or False Negative Rate or FNR)~\cite{kairouz2015composition,wasserman2010statistical}. More precisely, $(\eps, \del)$-DP implies that for all tests $\phi$, the inequality $1 - \fnr_\phi \leq e^\eps \fpr_\phi + \del$ holds. The work in~\cite{dong2019gaussian} formalize this hypothesis testing interpretation as the following differential privacy definition.
\begin{definition}[$f$-differential privacy~\cite{dong2019gaussian}]
  A mechanism $\M$ is $f$-\emph{differentially private} ($f$-DP) if for all $\x,\x' \in \X$, with the output distributions $\M(\x), \M(\x')$ denoted as $P, Q$ respectively, the following condition holds:
  \begin{equation}
    \forall \fpr \in [0,1] \ :\  T_{P \Vert Q}(\fpr) \geq f(\fpr),
  \end{equation}
  where $T_{P\Vert Q}$ is the \emph{trade-off function} defined as
  \begin{equation}
    T_{P\Vert Q}(\fpr) \eqdef \inf_{\phi:\Y\rightarrow\{0,1\}} \{\fnr_\phi : \fpr_\phi \leq \fpr\},
  \end{equation}
  with $\fpr_\phi \eqdef \expec{P}{\phi}$, and $\fnr_\phi \eqdef 1 - \expec{Q}{\phi}$.
\end{definition}
We say that a trade-off curve $f_1$ \emph{is dominated by} $f_2$ (denoted $f_1 \preccurlyeq f_2$), if:
\begin{equation}
  \forall \fpr \in [0,1] \ : \ f_1(\fpr) \geq f_2(\fpr).
\end{equation}

An $(\eps, \del)$-DP guarantee can be equivalently stated in terms of the $f$-DP guarantee.
\begin{proposition}[$(\eps, \del)$-DP to $f$-DP~\cite{dong2019gaussian}]
  \label{prop:dp_to_f}
  A mechanism is $(\eps, \del)$-DP if and only if it is $f_{\eps,\del}$-DP, where 
  \begin{equation}
    f_{\eps,\del}(\fpr) \eqdef \max\{0, 1 - \del - e^\eps \fpr, e^{-\eps}(1- \del - \fpr)\}.
\end{equation}
\end{proposition}
And, an $f$-DP guarantee and the curve comprising of all $(\eps, \del)$-DP guarantees are linked via \emph{convex-conjugatation}.
\begin{proposition}[$f$-DP to $(\eps, \del)$-DP~\cite{dong2019gaussian}]
  \label{prop:f_to_dp}
  A mechanism is $f$-DP if and only if it is $(\eps, \del_f(\eps))$-DP for all $\eps \geq 0$, where 
  \begin{equation}
    \del_f(\eps) \eqdef \sup_{\fpr \in [0,1]} 1 - \fpr \cdot e^\eps - f(\fpr).
  \end{equation}
\end{proposition}
More importantly, dominance in trade-off functions $f$ directly translates to dominance in their $\del_f(\eps)$ curves.
\begin{proposition}[{{\cite[Lemma B.2]{dong2019gaussian}}}]
  \label{prop:f_to_dp_dom}
  If trade-off functions $f_1, f_2$ satisfy $f_1 \preccurlyeq f_2$, then $\del_f(\eps) \leq \del_f(\eps)$ for all $\eps \geq 0$.
\end{proposition}

A DP guarantee is an upper bounds that is established \emph{in theory, typically by means of a pen-and-paper proof}. However, these guarantees can fail in practice \emph{if the analysis is incorrect or there are bugs in the implementation}. Unfortunately, such mistakes are not uncommon and many papers are written to exclusively point out these mistakes~\cite{chen2015privacy,lyu2016understanding,mironov2012significance,casacuberta2022widespread,lebeda2024avoiding}. 
Privacy auditing techniques provide an \emph{empirically lower bound} on the DP parameters by testing the implementation, which can help in detecting bugs or analytical errors when the lower and upper bounds contradict each other, or help assess the tightness of the upper-bounds otherwise~\cite{lokna2023group,steinke2024privacy,ding2018detecting,bichsel2021dp,mahloujifar2024auditing,zanella2023bayesian}.

In the following definition, we introduce the workflow for auditing DP adopted in this paper. Our framing here is derived from a compilation of existing auditing techniques, taking inspiration from~\cite{ding2018detecting,bichsel2021dp,bichsel2021dp,steinke2024privacy}, and formalizes the high-level principle for estimating an empirical lower bound on the privacy description of a mechanism.
\begin{definition}[$f$-DP Auditor]
  \label{dfn:f_dp_auditor}
  
  An $f$-DP auditor $(\A, \Rej_f)$ is a hypothesis test to determine whether the null-hypothesis ``mechanism $\M$ is $f$-DP'' holds or not.
  It either \emph{REJECT}s the null-hypothesis with a p-value at most $\sig$ or \emph{FAILS TO REJECT} it, as follows:
  \begin{enumerate}
    \item Auditor executes $\M$ in a controlled environment to collect certain statistics $\A(\M) \in \OO$.
    \item Auditor then \emph{REJECTS} if $\A(\M)$ lies in the rejection set $\Rej_f(\sig) \subset \OO$ of extreme values that rarely occur (probability less than $\sig$) under $f$-DP. Otherwise, the auditor \emph{FAILS TO REJECT}.
  \end{enumerate}
  Formally, an $f$-DP auditor $(\A, \Rej_f)$ is such that for all mechanisms $\M: \X \rightarrow \Y$ and significance levels $\sig \in [0,1]$,
  \begin{equation}
    \label{eqn:fdp_auditor}
    \M \text{ is } f\text{-DP} \implies \prob{}{\A(\M) \in \Rej_f(\sig)} \leq \sig.
  \end{equation}
\end{definition}
In other words, an \emph{$f$-DP auditor $(\A, \Rej_f)$ has only a small probability ($\leq\sig$) of incorrectly flagging an $f$-DP mechanism $\M$ as violating \( f \)-DP}. Crutially, this false-rejection probability $\sig$ can be controlled by adjusting the size of the rejection set $\Rej_f(\sig)$. 
Moreover, since $(\eps, \del)$-DP is equivalent to $f_{\eps, \del}$-DP (cf.~\autoref{prop:dp_to_f}), an $f$-DP auditor can also be used to audit $(\eps, \del)$-DP guarantees. 

\autoref{dfn:f_dp_auditor} provides a unifying perspective on the DP auditing techniques in the literature, which differ only in the attack statistic they evaluate ($\A(\M)$) and in how they construct the corresponding rejection region ($\Rej_f(\sig)$).

Steinke et al.~\cite{steinke2024privacy} show that a DP auditor can be leveraged to compute a high-confidence lower bound on the true DP characteristic of a mechanism $\M$. The following proposition restates the statement in~\cite[Lemma 4.7]{steinke2024privacy} but molded for $f$-DP auditing instead of $(\eps, \del)$-DP auditing.
\begin{proposition}[{{\cite[Lemma 4.7]{steinke2024privacy}}}]

  \label{lem:steinke_test_to_est}
  For each mechanism $\M$, let $f_\M \in \Fclass$ be the true $f$-DP guarantee it satisfies within a trade-off function family $\Fclass$. Let $(\A,\Rej)$ be an $f$-DP auditor for all $f \in \Fclass$. Further suppose that if $f_1 \preccurlyeq f_2$ within $\Fclass$, then the rejection set $\Rej_{f_1}(\sig) \supset \Rej_{f_2}(\sig)$. Then, for all $\M$ and all $\sig > 0$, 
  \begin{equation}
    \label{eq:steinke_test_to_est}
    \prob{}{f_\M \succcurlyeq  f^\mathrm{lb}} \geq 1 - \sig,
  \end{equation}
  where the trade-off curve $f^\mathrm{lb}$ is defined as
  \begin{equation}
    f^\mathrm{lb} \eqdef \sup \{f \in \Fclass : \A(\M) \in \Rej_f(\sig) \}.
  \end{equation}
\end{proposition}

The idea expressed in~\autoref{lem:steinke_test_to_est} turns a hypothesis test for $f$-DP, for which we can control the p-value $\sig$ by adjusting the size of rejection set $\Rej_f(\sig)$ (see Eqn.~\eqref{eqn:fdp_auditor}), into an \emph{$f$-DP lower bound estimator} that searches a class $\Fclass$ for a maximal trade-off curve $f^\mathrm{lb}$ that has at least $1 - \sig$ probability of being rejected by the auditor. In other words, the trade-off curve $f^\mathrm{lb}$ serves as a high-probability lower bound on the unknown $f_\M$-DP guarantee that $\M$ truly satisfies.

The choice of the trade-off family $\Fclass$ determines the shape of the lower bound $f^\mathrm{lb}$ that we get on auditing. If we do not know anything about the mechanism $\M$, it is reasonable to use the family $\Fclass_\eps \eqdef \{f_{\eps,\del} : \eps \geq 0 \}$ or $\Fclass_\del \eqdef \{f_{\eps, \del} : \del \in [0,1]\}$ denoting the trade-off curves corresponding to $(\eps,\del)$-DP for fixed $\eps$ or fixed $\del$ respectively. However, Nasr et al.~\cite{nasr2023tight} points out that typical trade-off curves of many mechanisms are considerably distinct from the $\Fclass_\eps$ or $\Fclass_\del$ family, which causes lower bound estimates from these families to underestimate the true $f$-DP guarantee of such mechanisms. If we know what mechanism we are auditing, e.g. Gaussian or Laplace mechanisms, we can restrict the audit to the respective family of trade-off curves, such as Gaussian family $\Fclass_\gaus$ or Laplace family $\Fclass_\lap$, to get the tighter lower-bound estimate $f^\mathrm{lb}$ on the true $f$-DP guarantee $f_\M \in \Fclass$ that $\M$ satisfies. 
\begin{remark}
  \label{rem:gauss_lap_family}
  The Gaussian and Laplace family of trade-off curves are defined respectively as
  \begin{align}
    \Fclass_\gaus &\eqdef \{ T_{\mathcal{N}(0, 1) \Vert \mathcal{N}(\mu, 1)} \mid \mu \geq 0\}, \\
    \Fclass_\lap &\eqdef \{T_{\lap(0, 1) \Vert \lap(\mu, 1)} \mid \mu \geq 0\}.
  \end{align}
\end{remark}
The duality between trade-off curves $f(\fpr)$ curve and $\del_f(\eps)$-curve in~\autoref{prop:f_to_dp} and the dominance relationship in~\autoref{prop:f_to_dp_dom} converts an $f$-DP lower bound into an $(\eps, \del)$-DP lower bound.
\begin{equation}
  f_\M \succcurlyeq f^\mathrm{lb} \implies \forall \eps \geq 0 \ : \ \del_{f_\M}(\eps) \geq \del_{f^\mathrm{lb}}(\eps).
\end{equation}
Essentially, if we have a lower bound curve $f^\mathrm{lb}$ for the real $f$-DP guarantee of a mechanism $\M$, it means we can confidently say that $\M$ is \emph{not $(\eps, \del_{f^\mathrm{lb}}(\eps))$-DP for any $\eps \geq 0$}. Alternatively, we can say that $\M$ is \emph{not $(\eps_{f^\mathrm{lb}}(\del), \del)$-DP for any $\del \in [0,1]$}, where $\eps_{f^\mathrm{lb}}(\del)$ is the \emph{right inverse function} of $\del_{f^\mathrm{lb}}(\eps)$, defined as
\begin{equation}
 \eps_{f^\mathrm{lb}}(\del) \eqdef \inf \{\eps \geq 0 : \del_{f^\mathrm{lb}}(\eps) \leq \del \}.
\end{equation}

\subsubsection*{Bayesian DP Auditor}
For auditing a mechanism $\M: \X \rightarrow \Y$, we design a \emph{membership inference attack $\phi: \Y \rightarrow \{0, 1\}$} to infer whether $\M$ was ran on an input $\x_0$ or $\x_1$ and \emph{empirically estimate its FPR $\fpr$ and  $\fnr$}. This is done by running $\M$ multiple times on a random choice of $\x_0$ or $\x_1$ and using the attack $\phi$ to predict the choice made for each run, which gives us a confusion matrix $(\tn, \fp, \fn, \tp)$. Zanella-B\'eguelin et al.~\cite{zanella2023bayesian} note that if the real false positive rate is $\fpr$, then the number of false positives ($\fp$) observed across all $\mathrm{N} := \fp + \tn$ trials (where the true input was $\x_0$) follows a binomial distribution:
\begin{equation}
  \fp | \fpr, \mathrm{N} \sim \Bin(\mathrm{N}, \fpr).
\end{equation}
Similarly, if the real false negative rate is $\fnr$, then the number of false negatives ($\fn$) observed across all $\mathrm{P} := \fn + \tp$ trials (where the true input was $\x_1$) follows:
\begin{equation}
  \fn | \fnr, \mathrm{P} \sim \Bin(\mathrm{P}, \fnr).
\end{equation}
Therefore, by assuming the real $\fpr, \fnr$ to be distributed according to the Beta distribution $\fpr, \fnr \sim \Beta(\sfrac{1}{2}, \sfrac{1}{2})$, which is the \emph{non-informative conjugate prior} for the Binomial distributions above, the Bayes rule allows them to model the posterior distribution of $(\fpr, \fnr)$ given the attack statistics $(\tp, \fp, \fn, \tn)$ as follows:
\begin{align}
  \fpr | \fp, \tn &\sim \Beta(\sfrac{1}{2} + \fp, \sfrac{1}{2} + \tn), \\
  \fnr | \fn, \tp &\sim \Beta(\sfrac{1}{2} + \fn, \sfrac{1}{2} + \tp).
\end{align}
By definition, if $\M$ is $f$-DP then all membership attacks must have FPR $\fpr$ and FNR $\fnr$ that satisfy $f(\fpr) \leq \fnr \leq 1 - f(1-\fpr)$. 
Consequently, the probability $p(f)$ that $\M$ satisfies $f$-DP for a candidate trade-off curve $f$ on observing the attack statistics $\A(\M) \eqdef (\tp, \fp, \fn, \tn)$ is
\begin{equation}
  p(f) = \prob{\substack{\fpr| \fp, \tn \\ \fnr|\fn, \tp}}{f(\fpr) \leq \fnr \leq 1 - f(1-\fpr)}.
\end{equation}
So if we define our rejection set as
\begin{equation}
  \Rej_f(\sig) \eqdef \left\{(\tp, \fp, \fn, \tn) : p(f) \leq  \sig \right\},
\end{equation}
then for all significance level $\sig \in [0,1]$, 
\begin{equation}
  \M \text{ is $f$-DP } \implies \prob{}{\M(\A) \in \Rej_f(\sig)} \leq \sig.
\end{equation}
In other words, the $(\A, \Rej_f)$ defined above describes an $f$-DP auditor. \autoref{alg:zb_auditor} describes this auditing approach.

\begin{algorithm}[!h]
  \DontPrintSemicolon
  \caption{$(\eps, \del)$-DP Estimator~\cite{zanella2023bayesian}}\label{alg:zb_auditor}
  \KwData{Mechanism $\M$, trade-off family $\Fclass$, significance threshold $\sig$, target delta $\del$, number of mechanism runs $n$}
  \KwResult{$\eps^\mathrm{lb}$ s.t. $\prob{}{\M \text{ is not } (\eps^\mathrm{lb}, \del)\text{-DP}} \geq 1 - \sig$}
   initialize $\x_0, \x_1 \in \X$ and membership test $\phi$\;
   uniformly sample $S \sample \{0, 1\}^n$\;
   \For{$i$ in $1, \cdots, n$ }{
     $Y_i \leftarrow \M(\x_{S_i})$ \tcp*{run mechanism}
     $T_i \leftarrow \phi(Y_i)$ \tcp*{run membership test}
   }
   \tcc{compute stats $\A(\M) \eqdef (\tn, \fp, \fn, \tp)$}
   $\A(\M) \leftarrow \mathbf{ConfusionMatrix}(S, T)$\;
   \tcc{formulate rejection set $\Rej_f(\sig)$}
   $\Rej_f(\sig) \eqdef \left\{(\tp, \fp, \fn, \tn) : p(f) \leq  \sig \right\}$\;
   \tcc{find $f^\mathrm{lb}$ to REJECT with $\sig$ significance}
   $f^\mathrm{lb} \leftarrow \sup\{f \in \Fclass:  \A(\M) \in \Rej_f(\sig)\}$\;
   \tcc{find corresponding $\eps^\mathrm{lb}$ for $\delta$ to REJECT}
   $\del_{f^\mathrm{lb}}(\eps) \eqdef \sup_{\fpr \in [0,1]} 1 - \fpr \cdot e^\eps - f^\mathrm{lb}(\fpr)$\;
   $\eps^\mathrm{lb} \leftarrow \inf \{\eps \geq 0 :  \del_{f^\mathrm{lb}}(\eps) \leq \del \} $\;
   \textbf{Return:} $\eps^\mathrm{lb}$\;
\end{algorithm}

\end{document}